\documentclass[a4paper,12pt]{article}
\usepackage{graphicx} 
\usepackage{epstopdf} 
\usepackage{authblk}

\begin{document}

\title{Accelerating Science TRIZ inventive methodology in illustrations}

\author[1]{Elena Seraia\thanks{\texttt{elena.seraia@ndm.ox.ac.uk}}}
\author[2]{Andrei Seryi}
\affil[1]{Nuffield Department of Medicine, \protect\\Target Discovery Institute HTS, University of Oxford\vspace*{1pc}}
\affil[2]{Department of Physics and John Adams Institute, \protect\\University of Oxford}


\date{29 July 2016}

\maketitle

\begin{abstract}
Theory of Inventive Problem Solving (TRIZ) \cite{Altshuller-book} is a powerful tool widely used in engineering community. It is based on identification of a physical contradiction in a problem, and based on the corresponding pair of contradicting parameters selecting a few of suitable inventive principles, narrowing down the choice and leading to a much faster solution of a problem.  

It is remarkable that TRIZ methodology can also be applied to scientific disciplines. Many of TRIZ inventive principles can be {\it post factum} identified in various in scientific inventions and discoveries. However, additional inventive principles, more suitable for scientific disciplines, should be introduced and added to standard TRIZ, and some of the standard inventive principles need to be reformulated to be better applicable to science --- we call this extension  Accelerating Science TRIZ \cite{Seryi-book}.

In this short note we describe and illustrate the AS-TRIZ inventive principles via scientific examples, identifying AS-TRIZ inventive principles in discoveries and inventions originated from physics, biology, and other areas. 

This brief note, we believe, is yet one more step towards bringing TRIZ methodology closer into the scientific community.
\end{abstract}

\section*{Introduction}

TRIZ -- Theory of Inventive Problem Solving -- was developed by Genrikh Altshuller in the Soviet Union in the mid-20th century \cite{Altshuller-book}. The ideas which led to development of TRIZ emerged when the author was working in 1946 in patent office. In the next several decades after that Altshuller and his team analyzed many thousands of patents, trying to discover patterns to identify what makes a patent successful. Following his work in the patent office, between 1956 and 1985 he formulated TRIZ and, together with his team, developed it further. Since then, TRIZ has gradually become one of the most powerful tools in the industrial world. For example, in his 7~March 2013 contribution to the business magazine Forbes, ``What Makes Samsung Such An Innovative Company?'', Haydn Shaughnessy wrote\footnote{http://www.forbes.com/sites/haydnshaughnessy/2013/03/07/why-is-samsung-such-an-innovative-company/} 
that TRIZ ``became the bedrock of innovation at Samsung'', and that ``TRIZ is now an obligatory skill set if you want to advance within Samsung''.

The authors of TRIZ devised the following four cornerstones for the method: the same problems and solutions appear again and again but in different industries; there is a recognizable technological evolution path for all industries; innovative patents (which are about a quarter of the total) use science and engineering theories outside of their own area or industry; and an innovative patent uncovers and solves contradictions. In addition, the team created a detailed methodology, which employs tables of typical contradicting parameters and a wonderfully universal table of 40 inventive principles. The TRIZ method consists in finding a pair of contradicting parameters in a problem, which, using the TRIZ inventive tables, immediately leads to the selection of only a few suitable inventive principles that narrow down the choice and result in a faster solution to a problem.

TRIZ method of inventiveness, although created originally for engineering, is universal and can also be applied to science. Indeed, TRIZ methodology provides another way to look at the world --- combined with science it creates a powerful and eye-opening amalgam of science and inventiveness. TRIZ is particularly helpful for building bridges of understanding between completely different scientific disciplines, and therefore its should be in particular useful to educational and research organizations that endeavor to break barriers between disciplines.

It is this ability to create bridges between different disciplines that inspired the authors to develop novel textbooks \cite{Seryi-book,Serye-book2} that connect physics of accelerators, lasers and plasma via the art and methodology of inventiveness. We will refer to the methods developed in these books further below. 

Still, experience shows that knowledge of TRIZ is nearly non-existent in the scientific departments of western universities. Moreover, it is not unusual to hear about unsuccessful attempts to introduce TRIZ into the graduate courses of universities’ science departments. Indeed, in many or most of these cases, the apparent reason for the failure is that the canonical version of TRIZ was introduced to science PhD students in the same way that TRIZ is taught to engineers in industrial companies. This may be a mistake, because science students are rightfully more critically minded and justifiably skeptical about overly prescriptive step-by-step methods. Indeed, a critically thinking scientist would immediately question the canonical number of 40 inventive principles, and would also note that identifying just a pair of contradicting parameters is a first-order approximation, and so on.

A more suitable approach to introduce TRIZ to graduate students, which takes into account the lessons learnt by our predecessors, could be different. Instead of teaching graduate students the ready-to-use methodology, it might be better to take them through the process of recreating parts of TRIZ by analyzing various inventions and discoveries from scientific disciplines, showing that the TRIZ inventive principles can be efficiently applied to science. In the process, additional inventive principles that are more suitable for scientific disciplines could be found and added to standard TRIZ, or the standard TRIZ principles would be adjusted. 

In our recent textbooks \cite{Seryi-book,Serye-book2}, we call this extension ``Accelerating Science (AS) TRIZ'', where ``accelerating'' refers not to charged particle accelerators, but instead highlights that TRIZ can help to boost various areas of science. The textbooks and the methodology itself is now starting to been used in science courses in Oxford, and intensively used in USPAS courses - US Particle Accelerator School, and demonstrated a boost in creativity of students.  

The illustrations of inventive principles presented in the following pages were in fact created for the recent USPAS-2016 course\footnote{http://uspas.fnal.gov/programs/2016/colorado/courses/unifying-physics.shtml} ``Unifying physics of accelerators, lasers and plasma''.  

In the following pages we will present illustrations of inventive principles with brief commentaries. We have decided for this note to keep the canonical number of 40 inventive principles, however we in some cases slightly renamed some of them or adjusted their description. 

The continuous process of adjustments the definitions and creating new examples should continue, and we encourage the readers to proactively participate in this process, as this would be the most efficient way to learn and master the TRIZ methodology.

\newpage
\section{Segmentation}

The inventive principle {\it segmentation} may involve dividing an object into independent parts, 
making an object easy to disassemble, or increasing the degree of fragmentation or segmentation.

\begin{figure}[!h]
\hspace*{-1pc}
\includegraphics[width=1.05\textwidth]{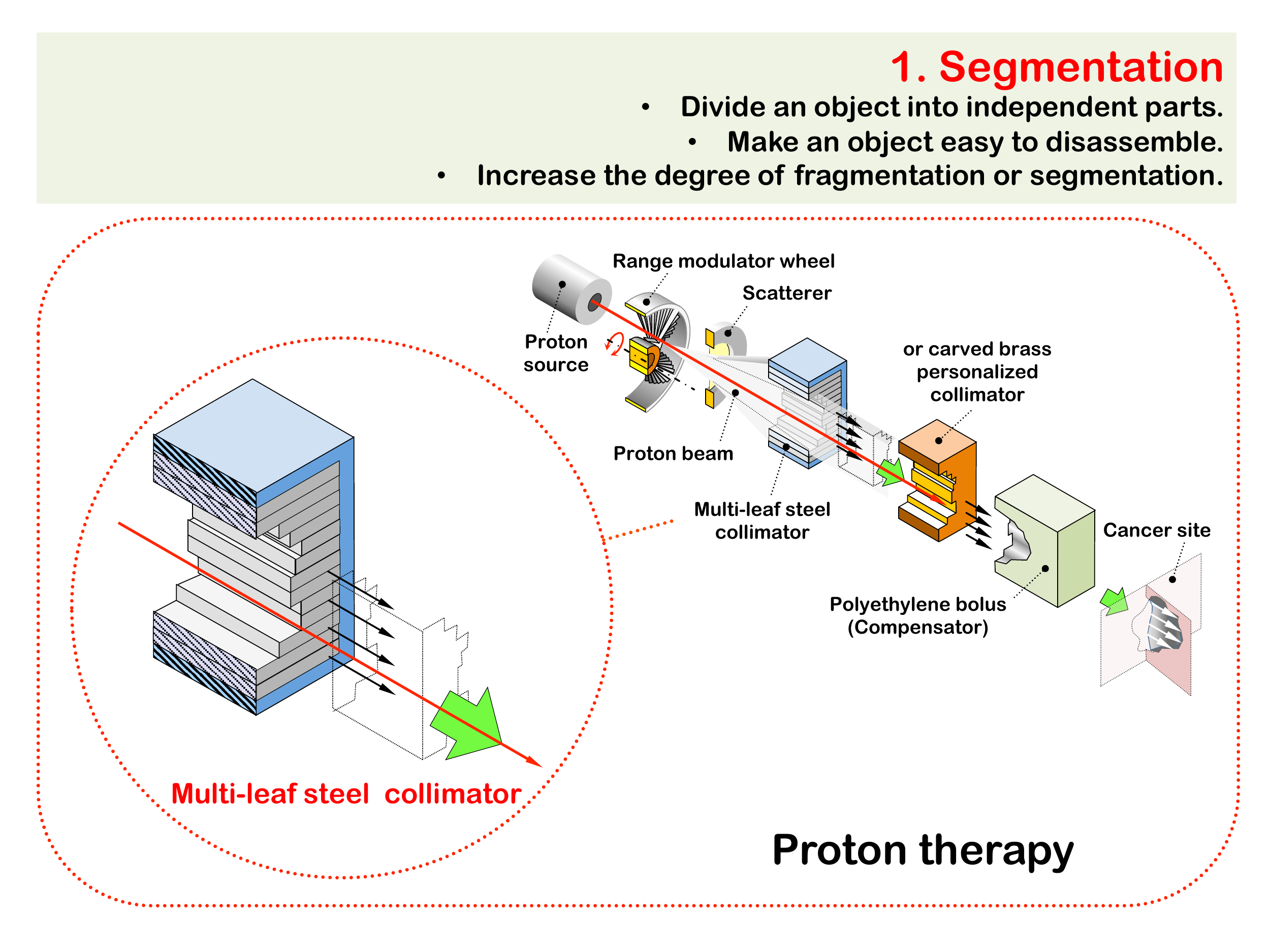}
\caption{Inventive principle ``Segmentation''.}
\label{lab01}
\end{figure}

An example we selected to illustrate this principle is a multi-leaf steel  collimator used in a beamline for particle therapy. This collimator is needed to shape the proton beam in such a way that it will correspond to the shape of the target (cancer site). Sometime, a solid personalized collimator used for these purposes, however it needs to be machined each time for specific case. An adjustable collimator, made from segmented steel leafs, make the treatment planning and delivery much more efficient.

\newpage
\section{Taking out}

The inventive principle {\it taking out} may involve separating an interfering part or property from an object or singling out the only necessary part (or \mbox{property}) of an object.

\begin{figure}[!h]
\hspace*{-1pc}
\includegraphics[width=1.05\textwidth]{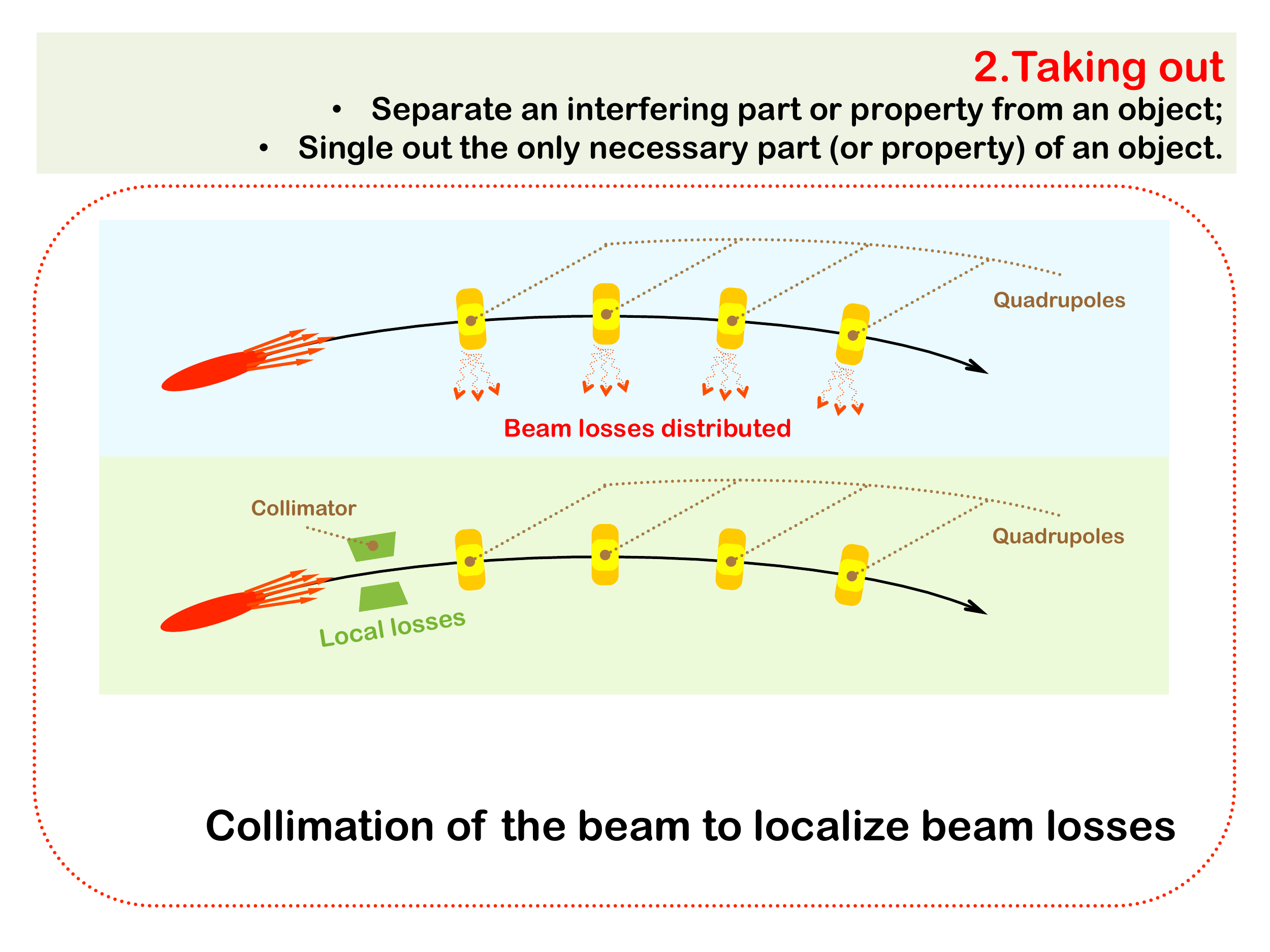}
\caption{Inventive principle ``Taking out''.}
\label{lab02}
\end{figure}

An example we selected to illustrate this principle is a collimator of the beam halo intended to  localize beam losses (which represent an interfering property in this case) in accelerators. 

The top picture above shows a beamline of an accelerator without a collimator. In this case particles from beam halo can be lost everywhere, creating problems associated with, for example, increased radiation in every location. 

Inserting a collimator in one location (bottom picture) would eliminate losses everywhere except the collimator itself, which itself can be treated specially, e.g. additional radiation shielding can be installed around the collimator. Therefore, by inserting the collimator we have separated the interfering property (losses) from the system.

\newpage
\section{Local quality}

The inventive principle {\it local quality} may involve changing an object's structure from uniform to non-uniform, changing an external environment (or external influence) from uniform to non-uniform, making each part of an object function in conditions most suitable for its operation, or 
making each part of an object fulfill a different and useful function. 

\begin{figure}[!h]
\hspace*{-1pc}
\includegraphics[width=1.05\textwidth]{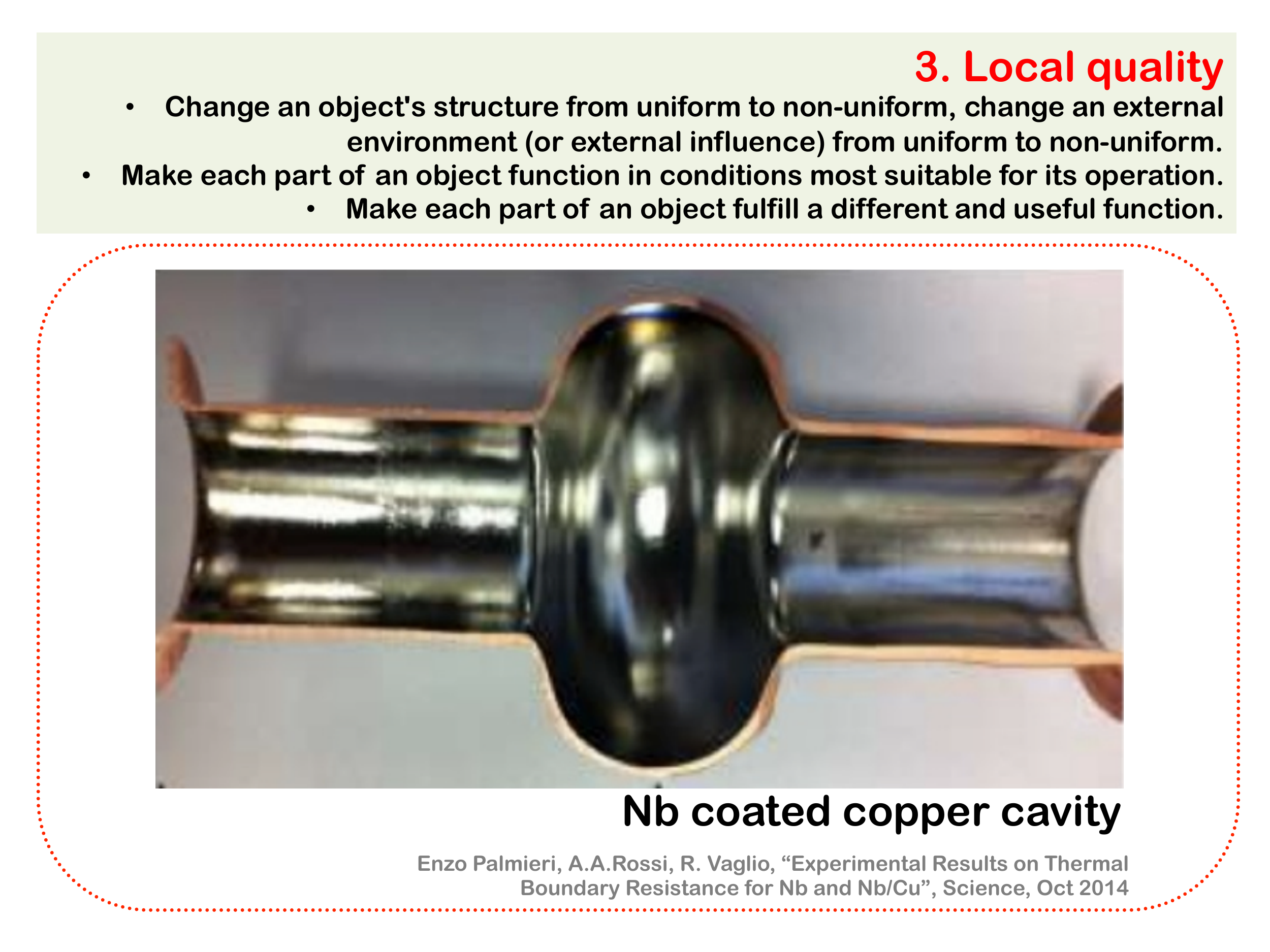}
\caption{Inventive principle ``Local quality''.}
\label{lab03}
\end{figure}

An example we selected to illustrate this principle is a superconducting resonator cavity, where the bulk of the cavity is made from copper, however the inner surface is covered by niobium. 

While typically the superconducting cavities are made entirely from niobium, which is an  expensive material, an arrangement as shown in the illustration would allow to considerably save on material cost of such cavities, provided, of course, that results of the ongoing studies\footnote {Enzo Palmieri, A.A. Rossi, R. Vaglio, {\it Experimental Results on Thermal Boundary Resistance for Nb and Nb/Cu}, Science, Oct 2014.} will be successful. 

\newpage
\section{Asymmetry}

The inventive principle {\it asymmetry} may involve changing the shape of an object from symmetrical to asymmetrical, or, if an object is asymmetrical, increasing the degree of its asymmetry.

\begin{figure}[!h]
\hspace*{-1pc}
\includegraphics[width=1.05\textwidth]{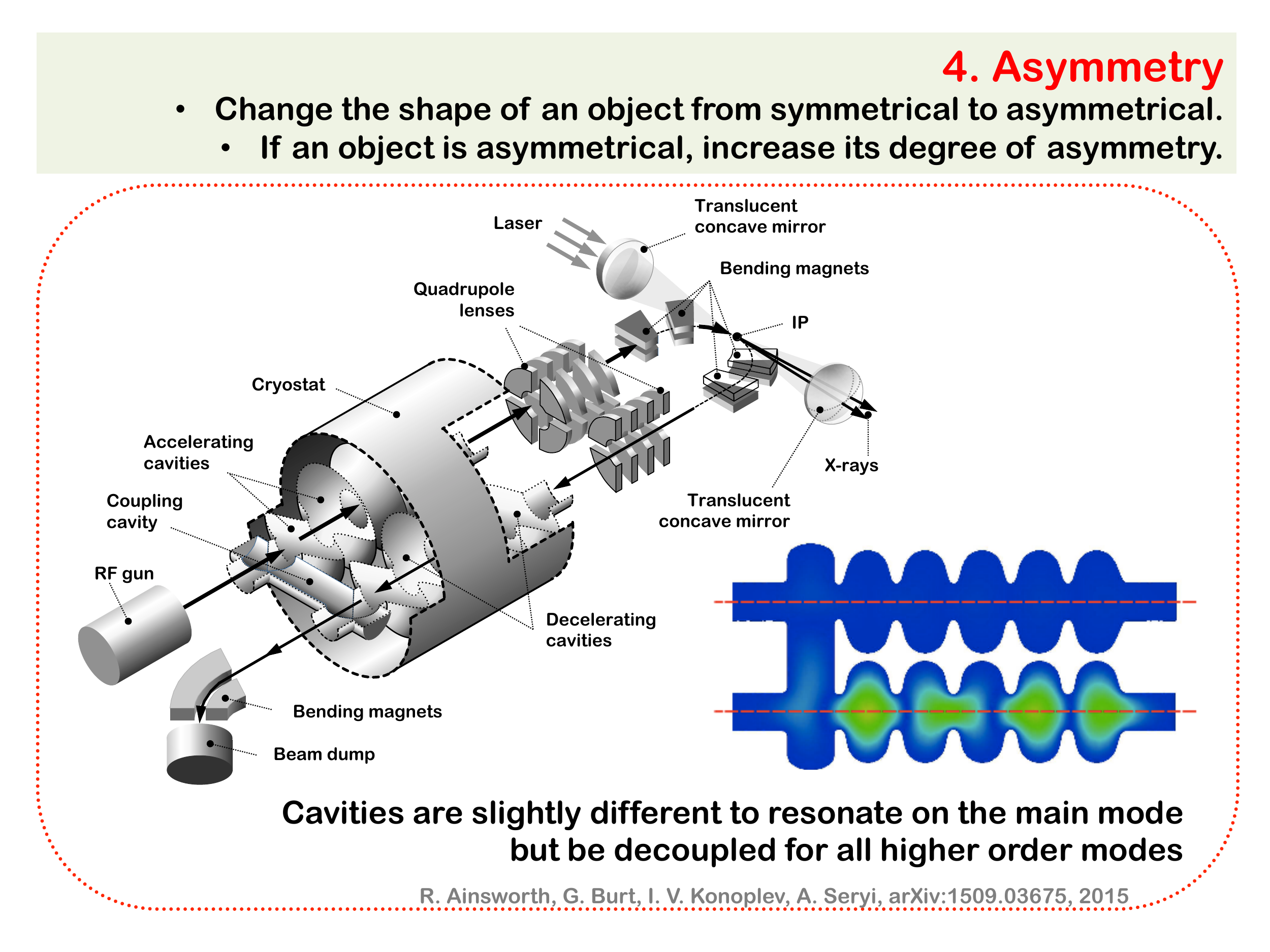}
\caption{Inventive principle ``Asymmetry''.}
\label{lab04}
\end{figure}

We illustrate this principle via asymmetrical design of the dual-axis coupled cavities in the compact energy-recovery based linac\footnote {R. Ainsworth, G. Burt, I. V. Konoplev, A. Seryi, {\it Asymmetric Dual Axis Energy Recovery Linac for Ultra-High Flux sources of coherent X-ray/THz radiation: Investigations Towards its Ultimate Performance}, arXiv:1509.03675, physics.acc-ph, Sep 2015.} shown above. 

In this linac an accelerated electron beam, after radiation generation, comes back to the decelerating part of the cavity, where the beam returns its energy to the system. In order to avoid instabilities of the beam which can be created in this system, all high order modes of the cavities need to be decoupled. This is achieved by introducing carefully designed asymmetry between every cell of the two cavities.

\newpage
\section{Merging}

The inventive principle {\it merging} may involve bringing closer together (or merging) identical or similar objects, assembling identical or similar parts to perform parallel operations, or
making operations contiguous or parallel; bringing them together in time.

\begin{figure}[!h]
\hspace*{-1pc}
\includegraphics[width=1.05\textwidth]{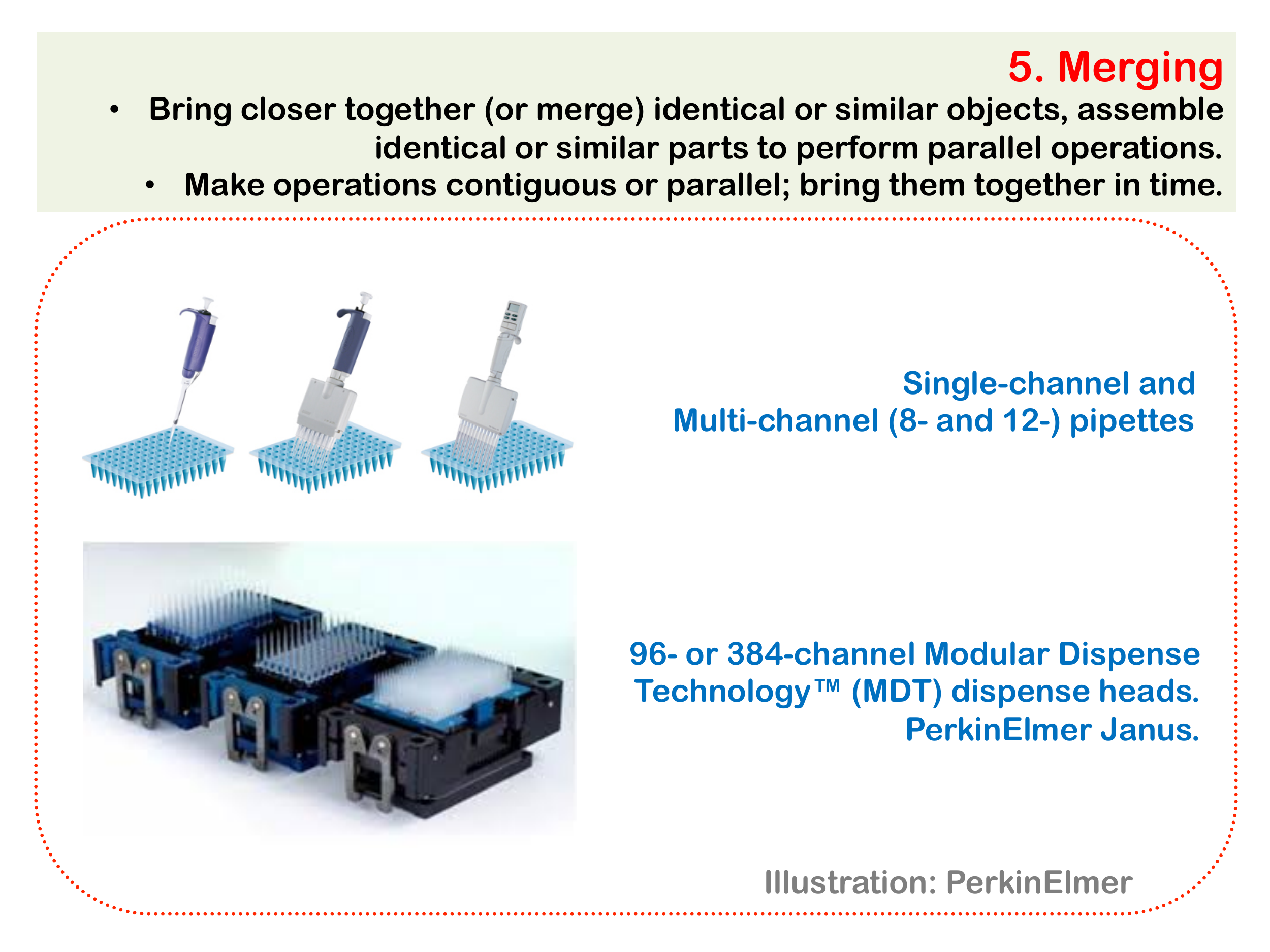}
\caption{Inventive principle ``Merging''.}
\label{lab05}
\end{figure}

An example we selected to illustrate this principle is multi-channel pipettes and modular dispensers that are now indispensable for biological studies where many samples, many genes, or many variations of drugs need to be studied and analyzed in parallel.

\newpage
\section{Universality}

The inventive principle {\it universality} may involve making a part or object perform multiple functions or eliminating the need for other parts.

\begin{figure}[!h]
\hspace*{-1pc}
\includegraphics[width=1.05\textwidth]{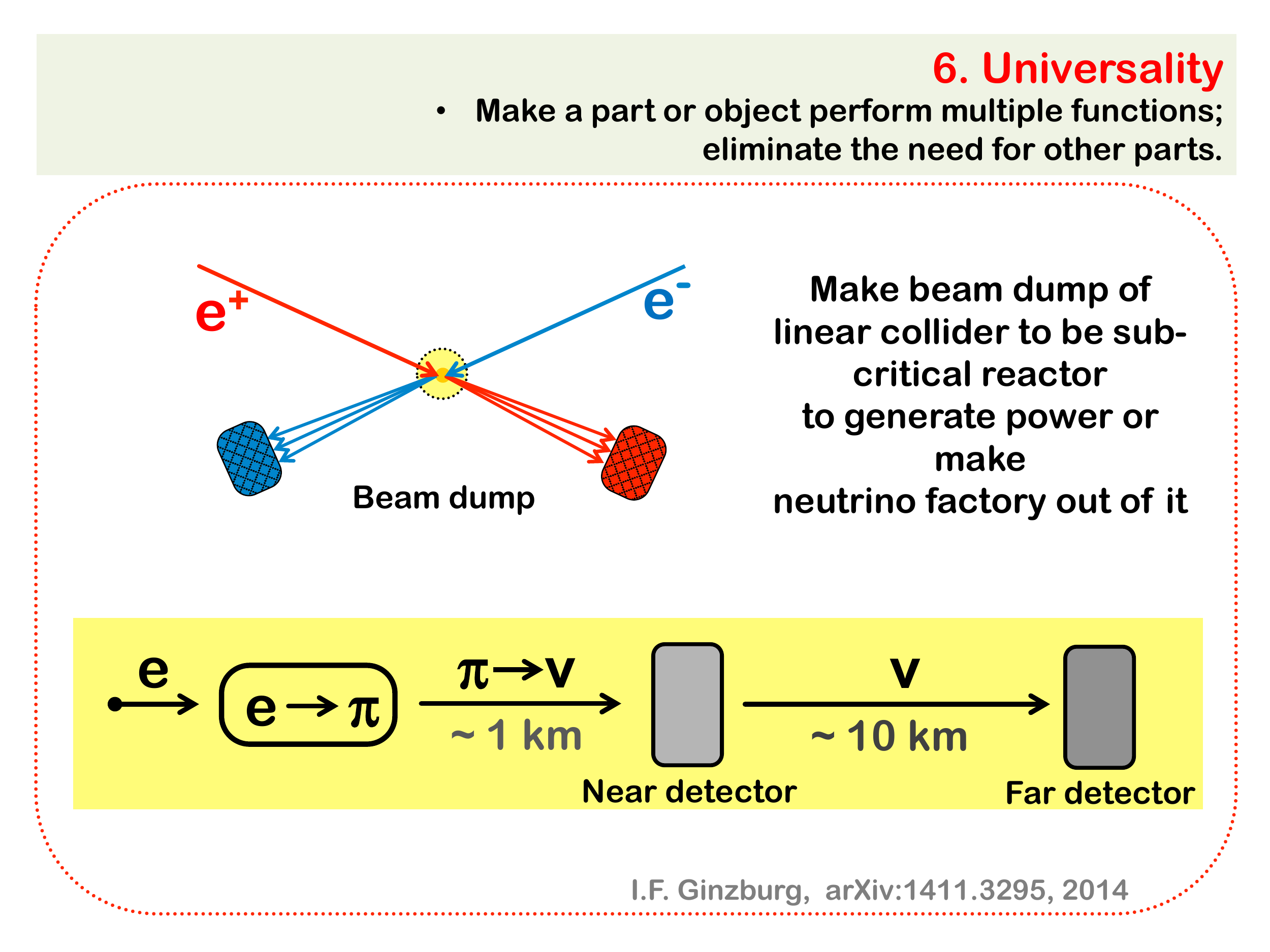}
\caption{Inventive principle ``Universality''.}
\label{lab06}
\end{figure}

An example we selected to illustrate this principle is the following peculiar design proposal for the beam dump of a linear collider. This beam dump needs to take and absorb, typically, 10 MW of CW power in the form of 250-500~GeV electron or positron beam. This energy is mostly waisted and goes to heat. A suggestion was made\footnote {I.F. Ginzburg, {\it Beam Dump problem and Neutrino Factory Based on a e+e− Linear Collider},  arXiv:1411.3295, physics.acc-ph, Oct 2014.} that this beam could in fact be used to either feed a sub-critical reactor to generate electric power or perhaps to make a neutrino factory out of it. Therefore, the beam dump of this design of a linear collider performs multiple functions and becomes universal. 

\newpage
\section{Nested doll}

The inventive principle {\it nested doll} may involve placing one object inside another, placing each object, in turn, inside the other, making one part pass through a cavity in the other.

\begin{figure}[!h]
\hspace*{-1pc}
\includegraphics[width=1.05\textwidth]{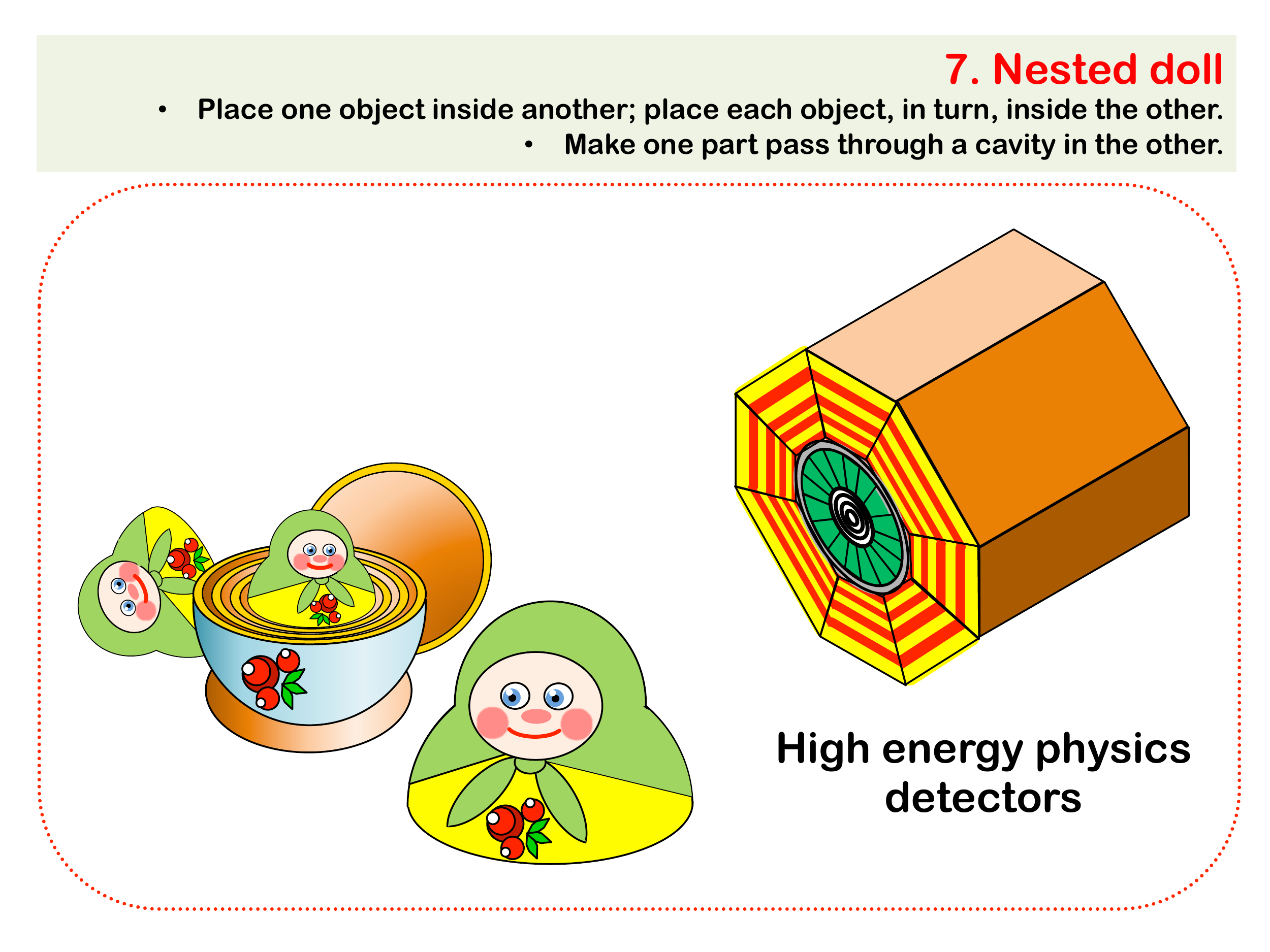}
\caption{Inventive principle ``Nested doll''.}
\label{lab07}
\end{figure}

An example we selected to illustrate this principle (which is also called inventive principle of {\it Russian dolls}) is the construction of a high-energy physics detector, where many different sub-detectors are inserted into one another, to enhance the accuracy of detecting elusive particles.  

\newpage
\section{Anti-force}

In standard TRIZ this principle is called ``anti-weight'', however for science applications it needs to be re-defined as ``anti-force'', since gravity often plays negligible role on particles or nano- and micro-objects science deals with, while electromagnetic forces can be much more important. 

The inventive principle {\it anti-force} may involve compensating for the force on an object, merging it with other objects that provide compensating force, etc., as explained in the figure below.  

\begin{figure}[!h]
\hspace*{-1pc}
\includegraphics[width=1.05\textwidth]{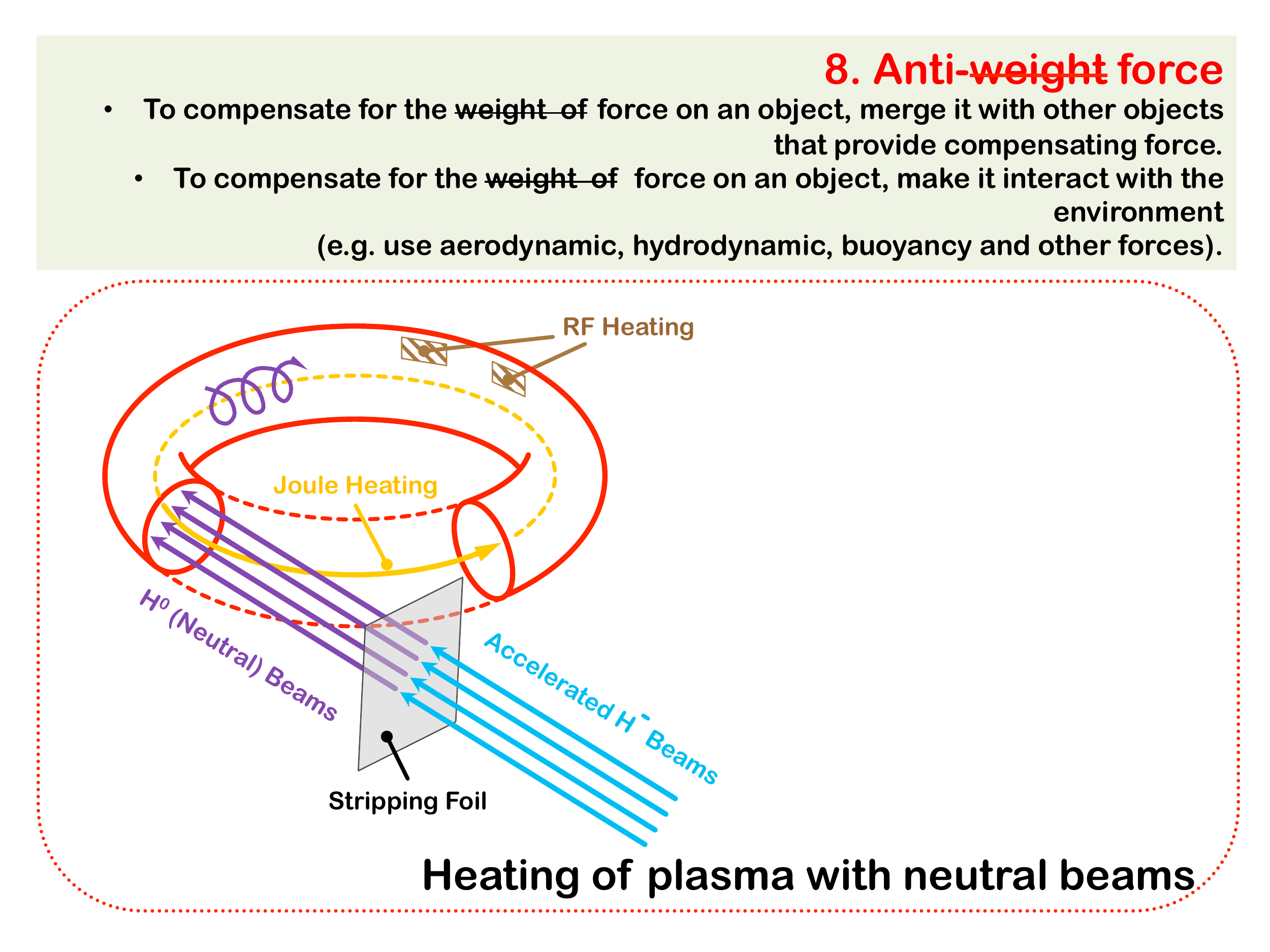}
\caption{Inventive principle AS-TRIZ ``Anti-force'', named ``Anti-weight'' in standard TRIZ.}
\label{lab08}
\end{figure}

We illustrate this principle with the heating system for plasma in Tokamak, where accelerated beam heats plasma. To avoid beam sensing the field of solenoid or plasma, the beam is made of neutral atoms, obtained by stripping electrons from the initial beam of  hydrogen negative ions.

\newpage
\section{Preliminary anti-action}

The inventive principle {\it preliminary anti-action} may involve replacing an action, which is known to produce both harmful and useful effects, with an anti-action to control those harmful effects. 

\begin{figure}[!h]
\hspace*{-1pc}
\includegraphics[width=1.05\textwidth]{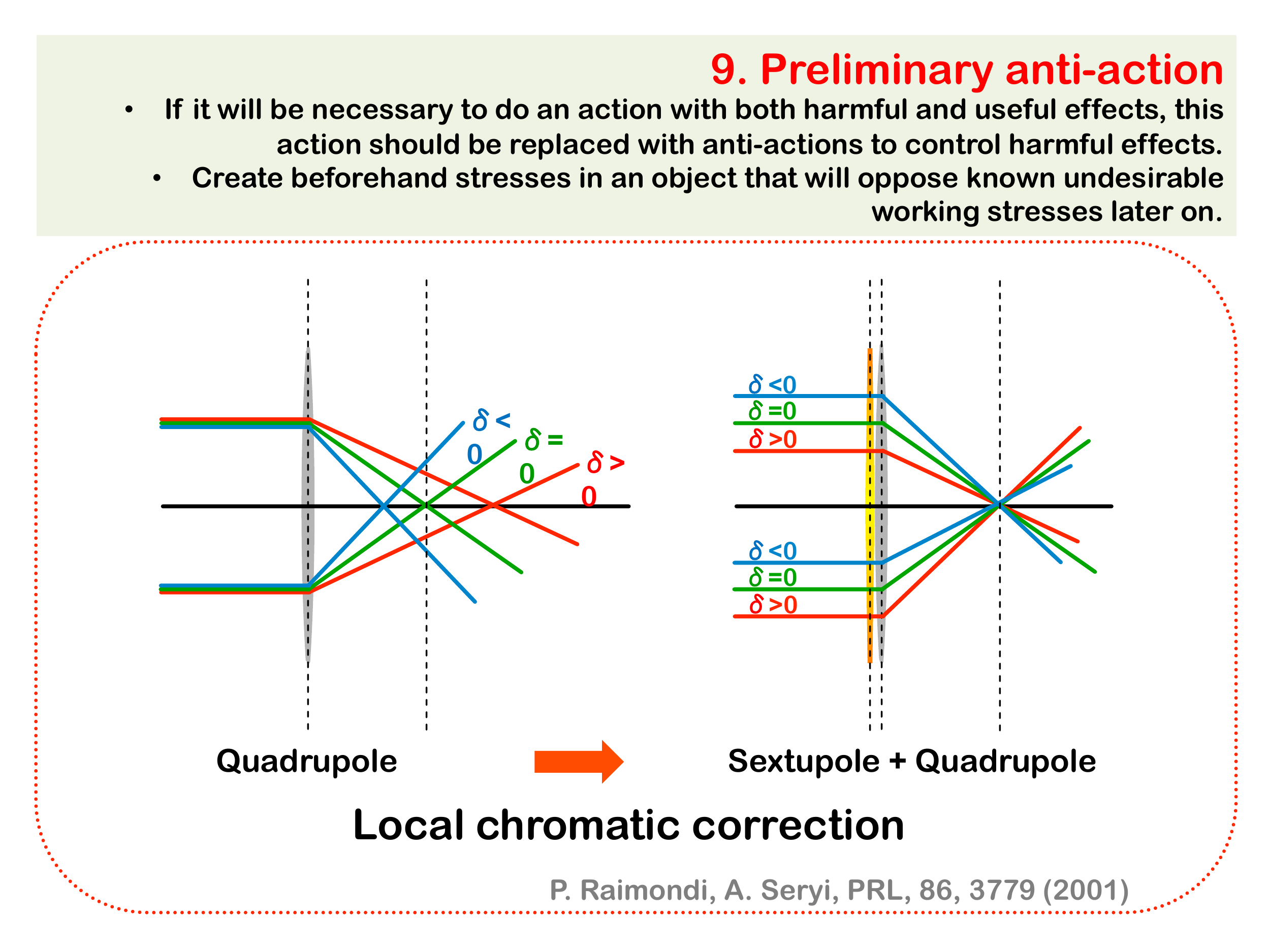}
\caption{Inventive principle ``Preliminary anti-action''.}
\label{lab09}
\end{figure}

An example we selected to illustrate this principle is a final focus with local chromatic correction\footnote {Pantaleo Raimondi and Andrei Seryi, {\it Novel Final Focus Design for Future Linear Colliders}, Phys. Rev. Lett., {\bf 86},3779, Apr 2001.}. Any strong focusing optics suffer from chromatic aberrations, as shown on the left part of the picture. Local chromatic correction involves dispersing the beam in energies {\it prior} it arrives to final lenses, and also inserting nonlinear magnet -- sextupole next to the final lens, which cancels the chromatic aberrations, thus acting against is. As a result of this preliminary anti-action the beam gets focused nicely into a tight spot as shown on the right picture.

\newpage
\section{Preliminary action}

The inventive principle {\it preliminary action} may involve performing, before it is needed, the required change of an object (either fully or partially), or pre-arranging objects such that they can come into action from the most convenient place and without losing time for their delivery. 

\begin{figure}[!h]
\hspace*{-1pc}
\includegraphics[width=1.05\textwidth]{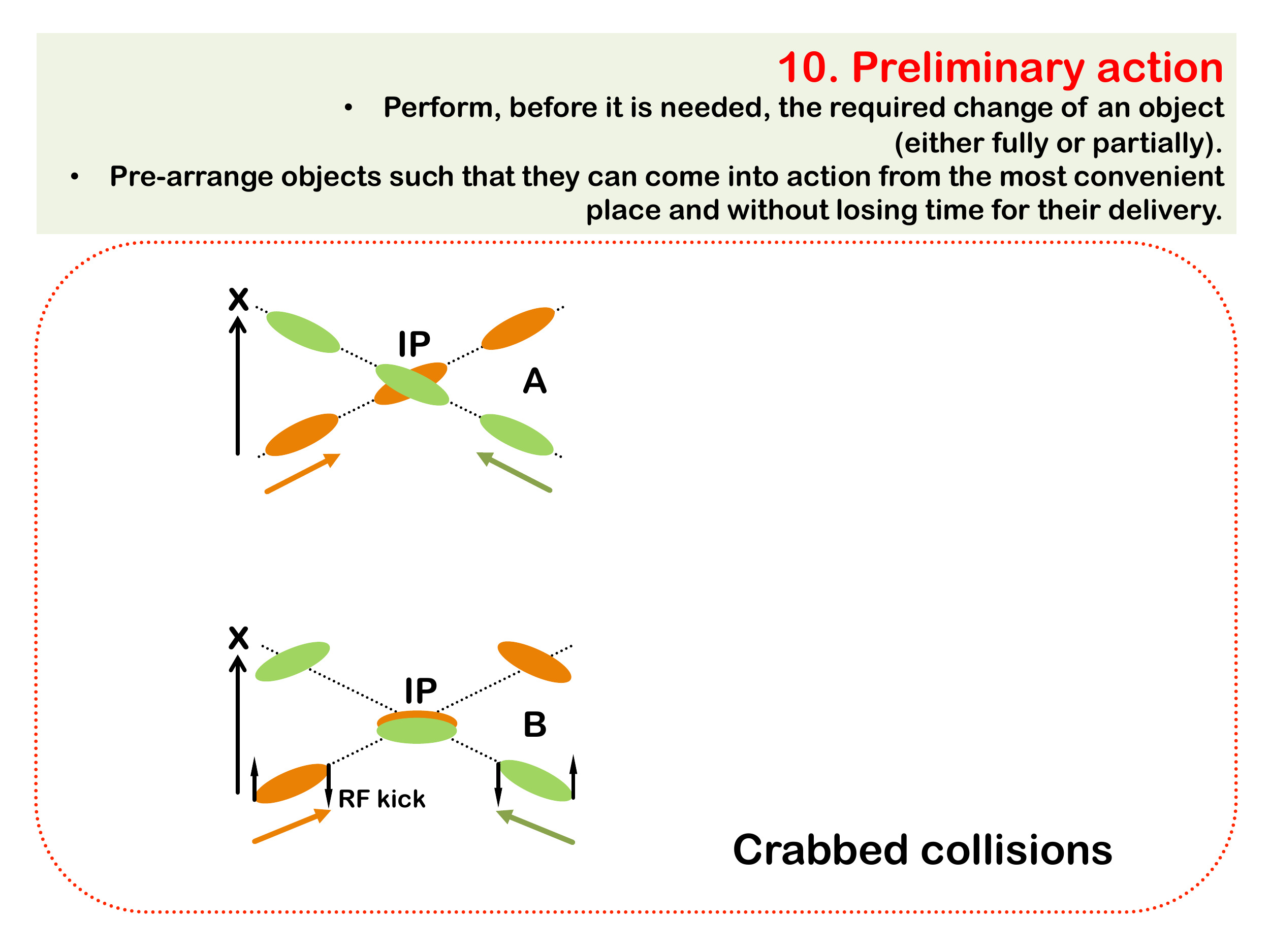}
\caption{Inventive principle ``Preliminary action''.}
\label{lab10}
\end{figure}

An example we selected to illustrate this principle is a crabbed collision. In a linear collider the electron and positron beams need to collide with a small crossing angle, as shown on the left side of the picture. However, their overlap during collision would then be incomplete and the luminosity would thus decrease. In order to prevent this loss, the beams, before collisions, can pass through a radio-frequency cavity, which would give to the beams a useful kick, in such a way that the head and tails of the beam receive kicks in different directions. The beams will therefore start to rotate and come to the collision point with a proper orientation, ensuring full overlap.

\newpage
\section{Beforehand cushioning}

The inventive principle {\it beforehand cushioning} may involve preparing emergency means beforehand to compensate for the relatively low reliability of an object.

\begin{figure}[!h]
\hspace*{-1pc}
\includegraphics[width=1.05\textwidth]{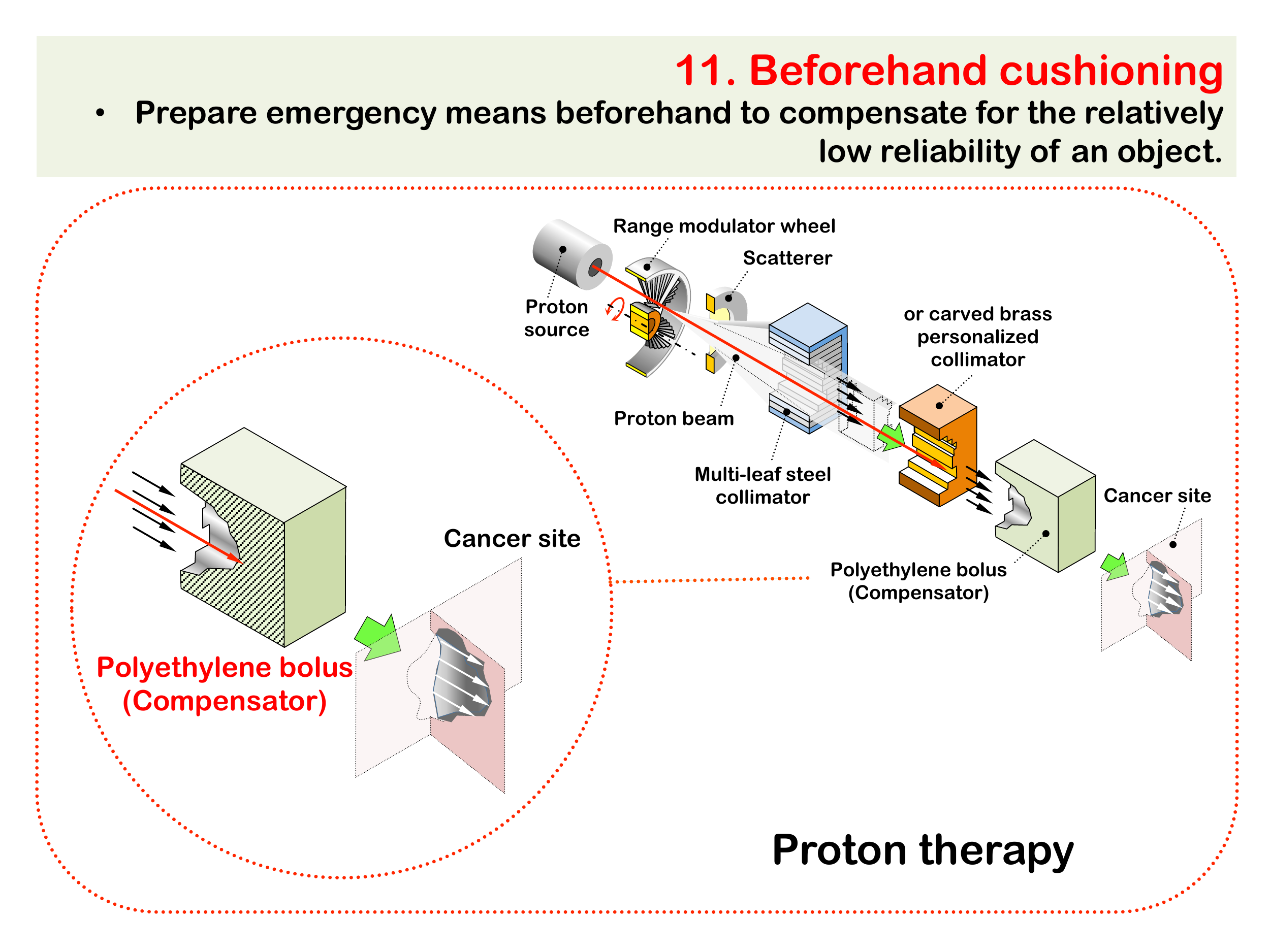}
\caption{Inventive principle ``Beforehand cushioning''.}
\label{lab11}
\end{figure}

An example we selected to illustrate this principle is a bolus (compensator) for the proton therapy beamline.  The thickness of the bolus is varied depending on location and 
therefore it will modify the energy of different parts of the proton beam, and correspondingly modify the penetration depth of the protons, matching the shape of the cancer site.  

The reader may argue that this example suit better the previous principle of preliminary action. If so, we would encourage the reader to suggest other examples, such as for example emergency kicker that would dump the beam in an accelerator in case of an accident, to prevent losses of the beam into precious superconducting magnets.

\newpage
\section{Equipotentiality}

The inventive principle {\it equipotentiality} may involve limiting position changes of an object in gravity or other potential field  (e.g. changing operating conditions to eliminate the need to raise or lower the objects in a gravity field).

\begin{figure}[!h]
\hspace*{-1pc}
\includegraphics[width=1.05\textwidth]{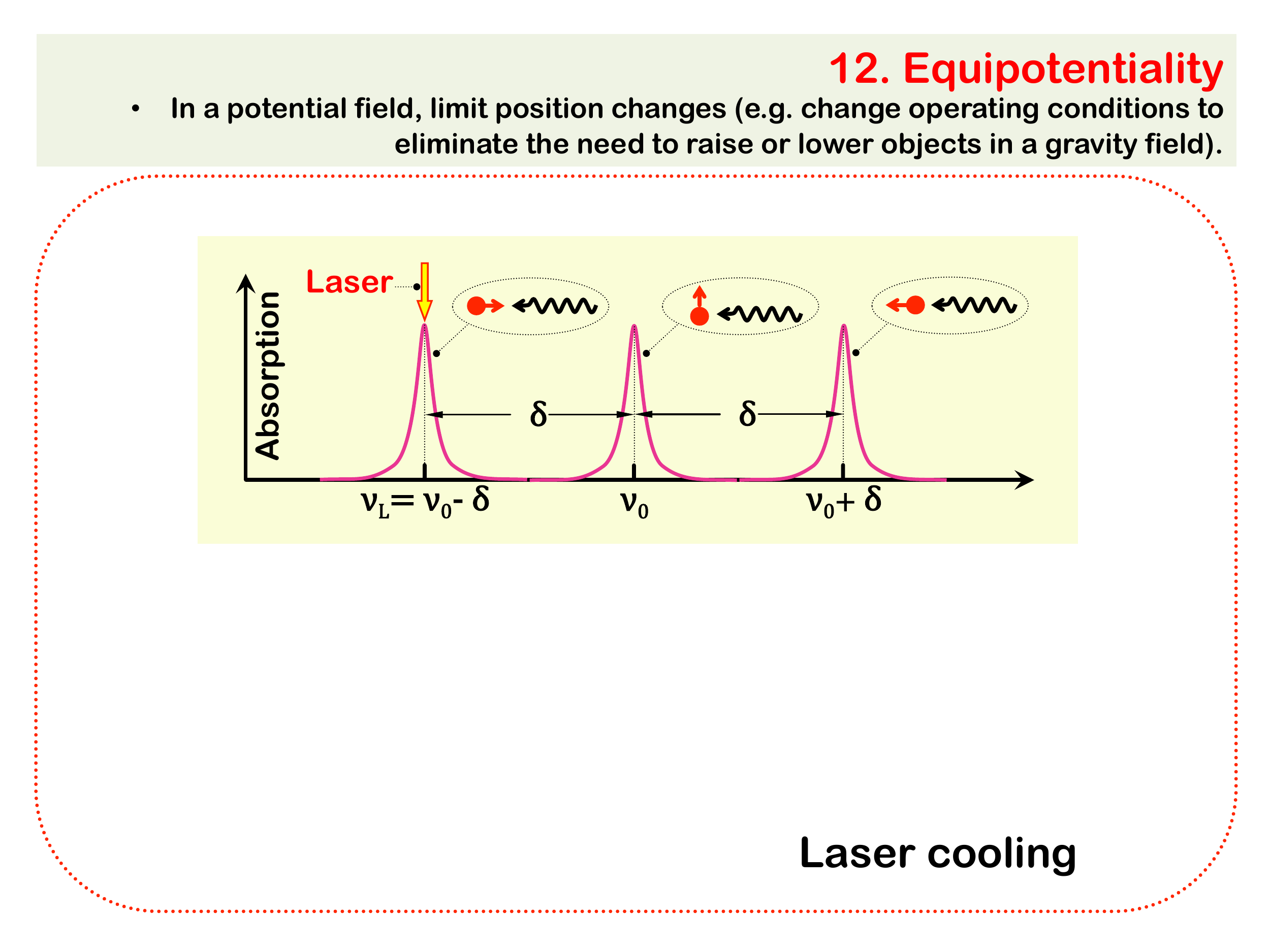}
\caption{Inventive principle ``Equipotentiality''.}
\label{lab12}
\end{figure}

An example we selected to illustrate this principle is a laser cooling, where interaction of laser with atom (excitation of the atom) occurs only when they are ``at the same potential'', i.e. when the velocity of the atom is such that due to Doppler shift the laser frequency corresponds to the energy of the atom excitation.  

\newpage
\section{The other way round}

The inventive principle {\it the other way round} (which can be also called principle of system and anti-system) may involve  inverting the action(s) used to solve the problem or turning 
the object (or process) ``upside down''. 

\begin{figure}[!h]
\hspace*{-1pc}
\includegraphics[width=1.05\textwidth]{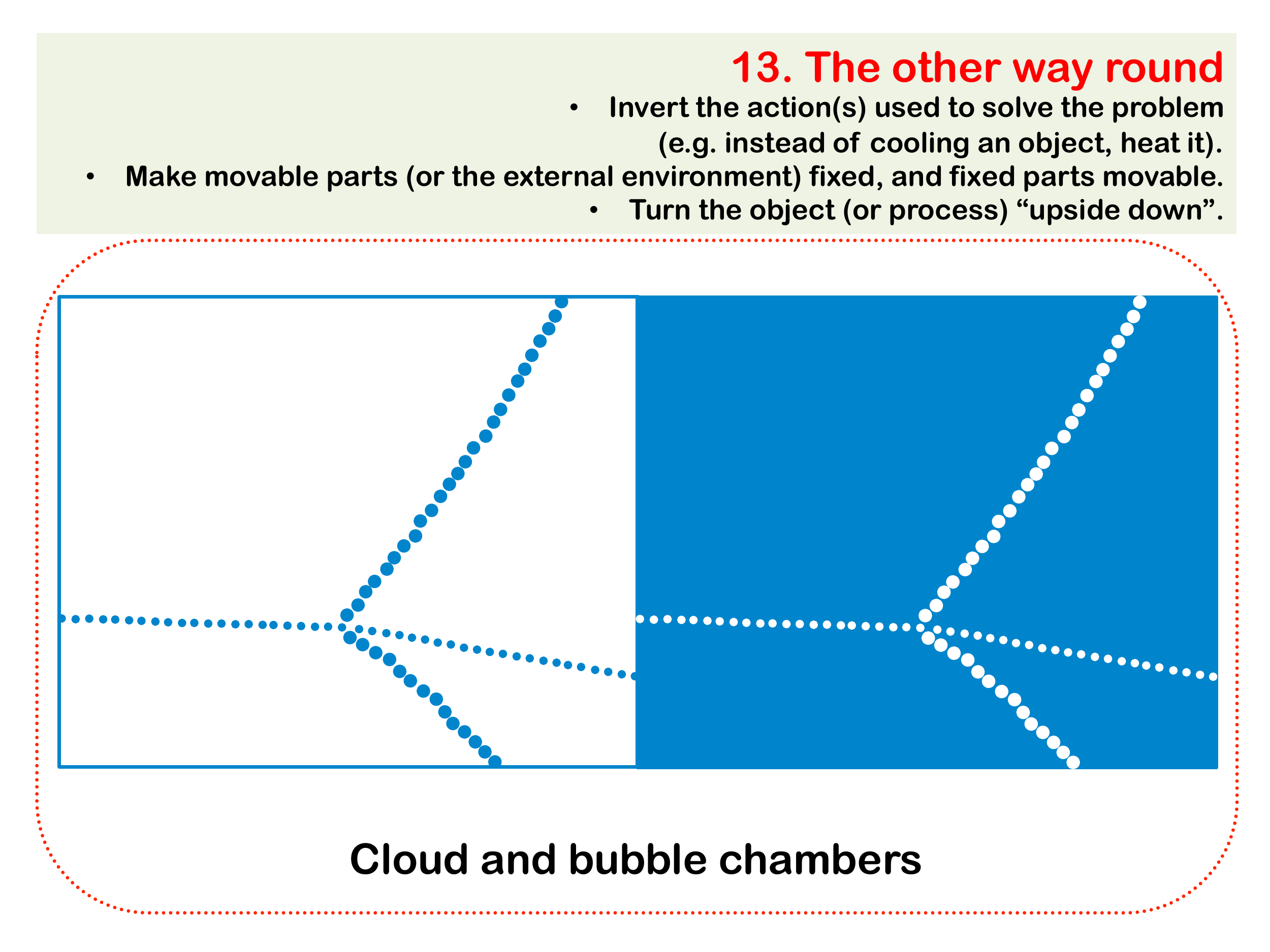}
\caption{Inventive principle ``The other way round''.}
\label{lab13}
\end{figure}

We illustrate this principle via consideration of the cloud and bubble chambers. TRIZ textbooks often cite Charles Wilson’s cloud chamber (invented in 1911) and Donald Glaser’s bubble chamber (invented in 1952) as examples of this principle ``system and anti-system''. 

Indeed, the cloud chamber works on the principle of bubbles of liquid created in gas, whereas the bubble chamber uses bubbles of gas created in liquid. If the TRIZ inventive principle of system/anti-system were applied, the invention of the bubble chamber would follow immediately and not almost half a century after the invention of the cloud chamber.

\newpage
\section{Spheroidality -- Curvature}

The inventive principle {\it spheroidality -- curvature} may involve using, instead of rectilinear parts, surfaces, or forms, curvilinear ones, moving from flat surfaces to spherical ones, etc.

\begin{figure}[!h]
\hspace*{-1pc}
\includegraphics[width=1.05\textwidth]{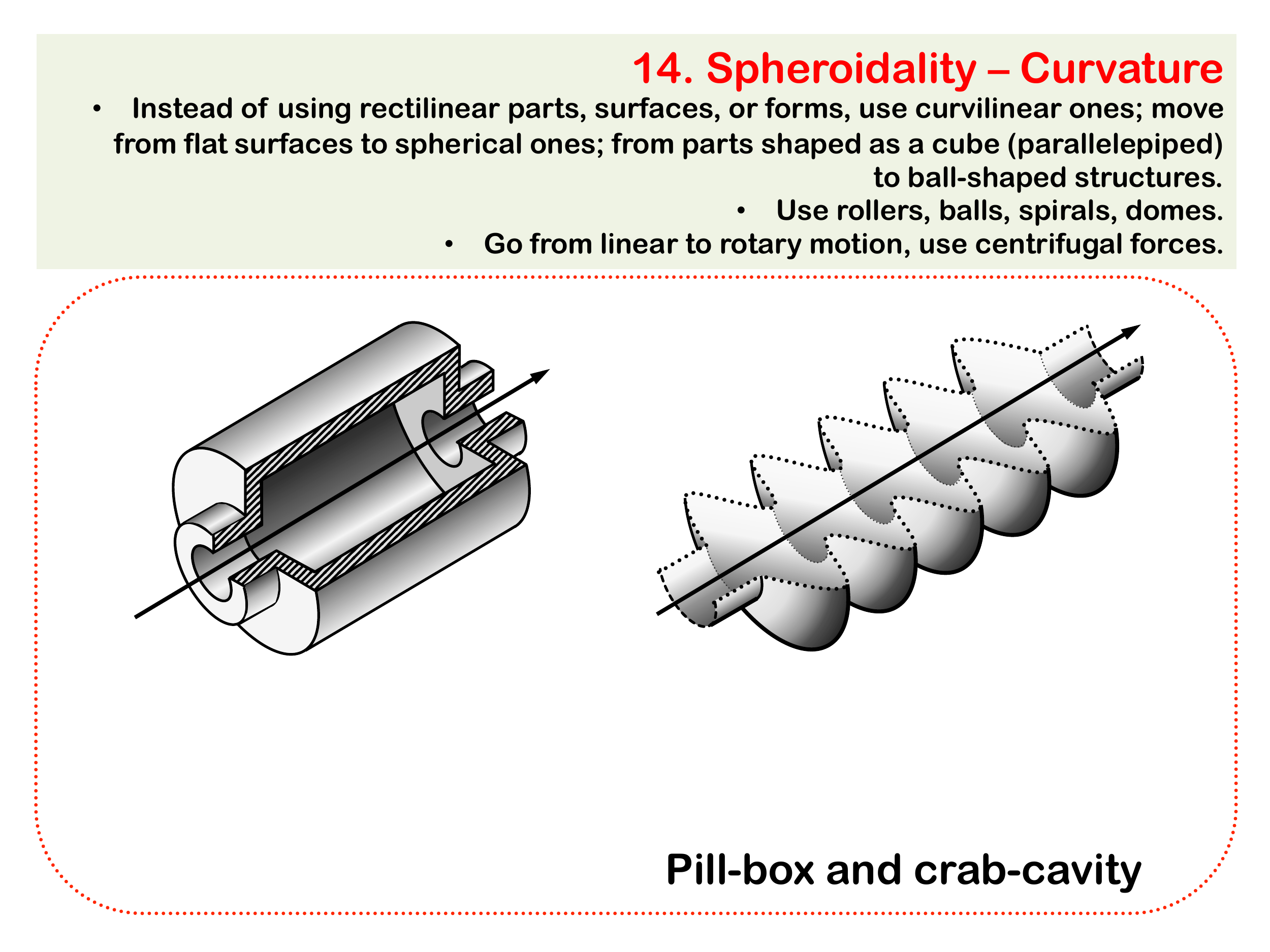}
\caption{Inventive principle ``Spheroidality -- Curvature''.}
\label{lab14}
\end{figure}

An example we selected to illustrate this principle is cavity resonator -- pill-box style as shown on the left side of the pictures, and elliptical cavity (where shapes are rounded and consist of various connected ellipses). A particular example shown on the right corresponds to the cavity which can produce crabbing kick mentioned in the principle 10, but it can also be any other similar cavity. Rounding the shapes of the resonator in such a way allows to achieve better and smooth distribution of fields and currents along the walls of the cavity and correspondingly higher fields generated by the cavity on beam axis.

\newpage
\section{Dynamics}

The inventive principle {\it dynamics} may involve  allowing (or designing) the characteristics of an object, external environment, or process to change to be optimal.  

\begin{figure}[!h]
\hspace*{-1pc}
\includegraphics[width=1.05\textwidth]{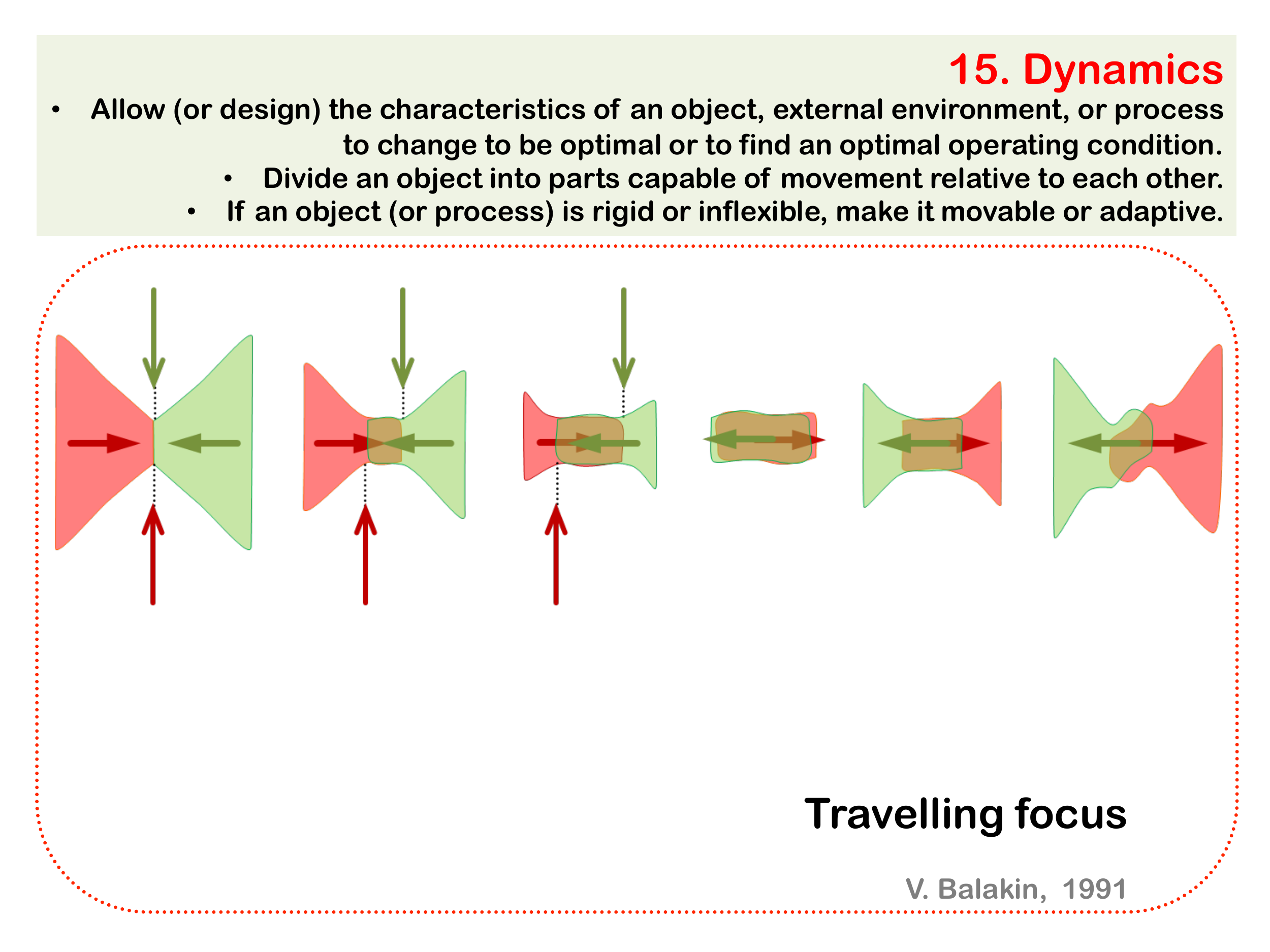}
\caption{Inventive principle ``Dynamics''.}
\label{lab15}
\end{figure}

An example we selected to illustrate this principle is a travelling focus idea\footnote {Balakin, V., {\it Travelling Focus Regime for Linear Collider VLEPP}, Proc. of 1991 IEEE PAC, p.3273, 1991.} intended to increase luminosity of linear colliders. The fields of the opposite beam during collision of e+ and e- beams can create an additional focusing which can help to squeeze the beams even tighter. However, for this additional focusing to work most optimally, the focal point for each beam needs to move during collision in such a way that it would coincide with the location of the head of the opposite beam.  The location of the focal point is shown by arrows of corresponding color. Such dynamic modification of the colliding beams would then give some increase of the luminosity.

\newpage
\section{Partial or excessive actions}

The inventive principle {\it partial or excessive actions} may involve, in case if 100\% of the effect is hard to achieve with a given solution or method, using ``slightly less'' or ``slightly more'' of the same method, to make the problem considerably easier to solve.

\begin{figure}[!h]
\hspace*{-1pc}
\includegraphics[width=1.05\textwidth]{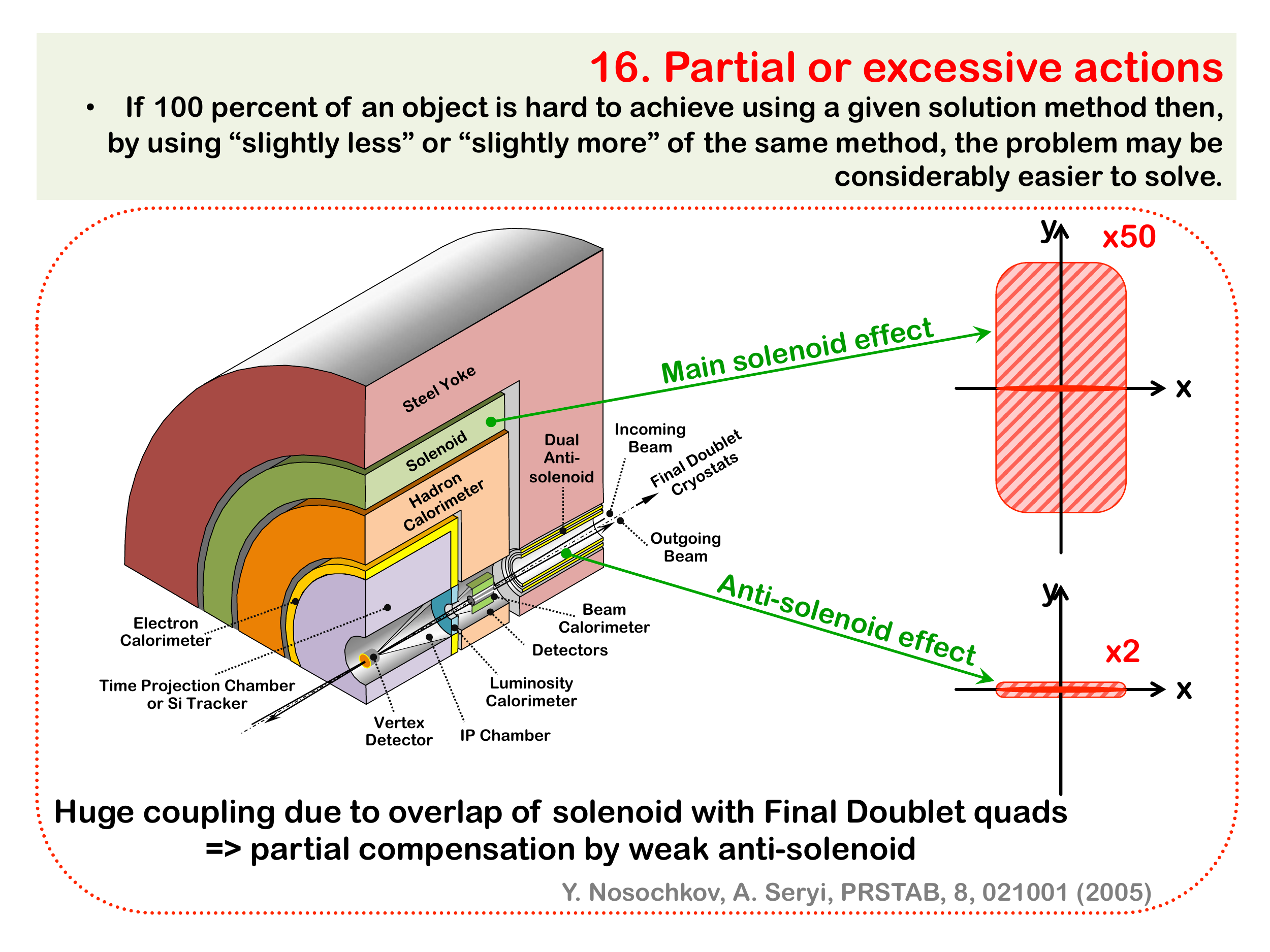}
\caption{Inventive principle ``Partial or excessive actions''.}
\label{lab16}
\end{figure}

An example we selected to illustrate this principle is the design concept of a weak antisolenoid\footnote {Y. Nosochkov, A. Seryi, PRSTAB, 8, 021001 , 2005.} intended for compensating the beam X-Y coupling effects in the interaction region of a linear collider. The anomalously large coupling effects arise due to overlap of the field of the main solenoid with the final focusing lenses. Properly adjusted weak anti-solenoid can compensate a large fraction of these detrimental effects, making the problem much easier to solve with upstream coupling correctors. 

\newpage
\section{Another dimension}

The inventive principle {\it another dimension} may involve  moving into an additional dimension or using a multi-story arrangement of objects instead of a single-story arrangement, or 
tilting or re-orienting the object, laying it on its side, etc. 

\begin{figure}[!h]
\hspace*{-1pc}
\includegraphics[width=1.05\textwidth]{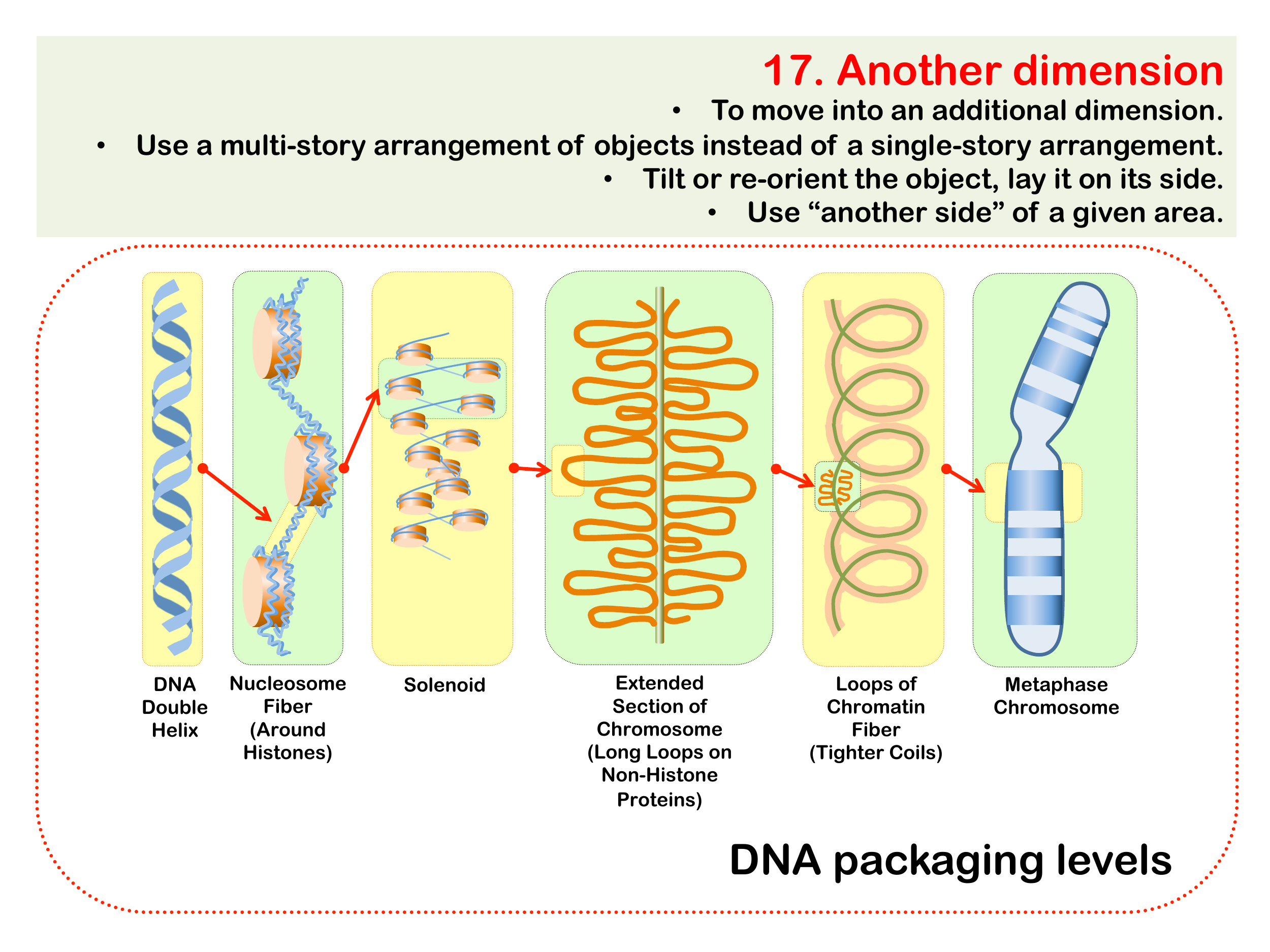}
\caption{Inventive principle ``Another dimension''.}
\label{lab17}
\end{figure}

An example we selected to illustrate this principle is something nature invented -- the DNA packaging mechanism.  The DNA molecules, if strengthened out, are quite long, around 2~mm. However, they are packaged in a cell in a compact way. In general, there are five main different packaging levels. First, by going into another, transverse, direction, the DNA molecules are packaged around histones. Next -- multiples of these assemblies are packaged in a solenoid-like way. This then further packaged with multiple loops, and so on.

\newpage
\section{Oscillations and resonances}

In standard TRIZ this method is called ``Mechanical vibration'', however for science applications this principle should better be re-defined as ``Oscillations and resonances''. 
The inventive principle {\it oscillations and resonances} may involve  causing an object to oscillate or vibrate, increasing its frequency (e.g. from microwave to optical), using an object's resonant frequency, etc. 

\begin{figure}[!h]
\hspace*{-1pc}
\includegraphics[width=1.05\textwidth]{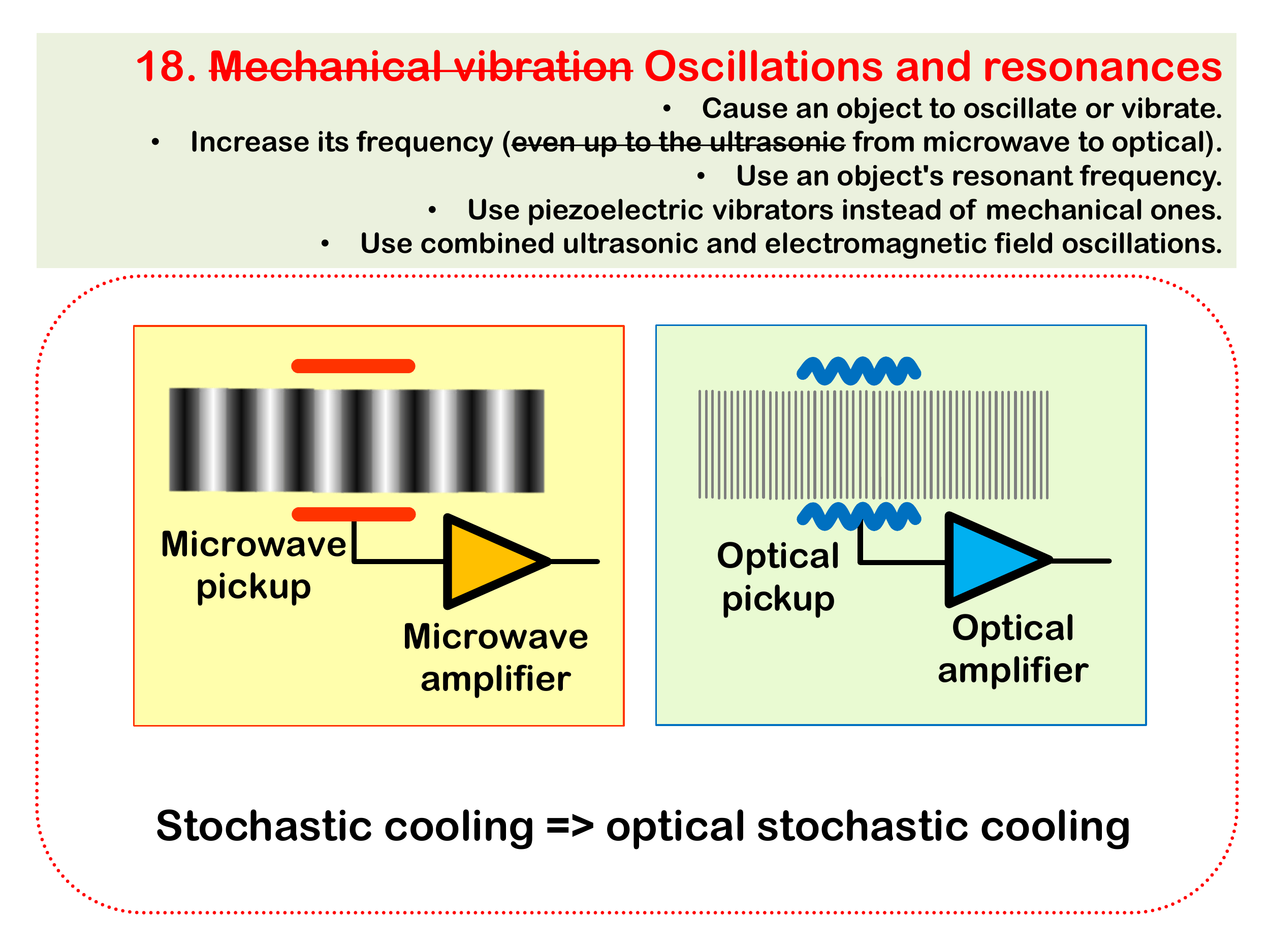}
\caption{Inventive principle AS-TRIZ ``Oscillations and resonances'', named ``Mechanical vibration'' in TRIZ .}
\label{lab18}
\end{figure}

We illustrate this principle with the design concept of optical stochastic cooling which represents further evolution of stochastic cooling. Both these methods are designed to decrease phase space volume of a beam of charged particle in accelerators. Stochastic cooling relies on microwave range of frequencies, for detection of particles and acting on them, while optical stochastic cooling relies on, correspondingly, optical frequencies.

\newpage
\section{Periodic action}

The inventive principle {\it periodic action} may involve, instead of continuous action, using periodic or pulsating actions. 

\begin{figure}[!h]
\hspace*{-1pc}
\includegraphics[width=1.05\textwidth]{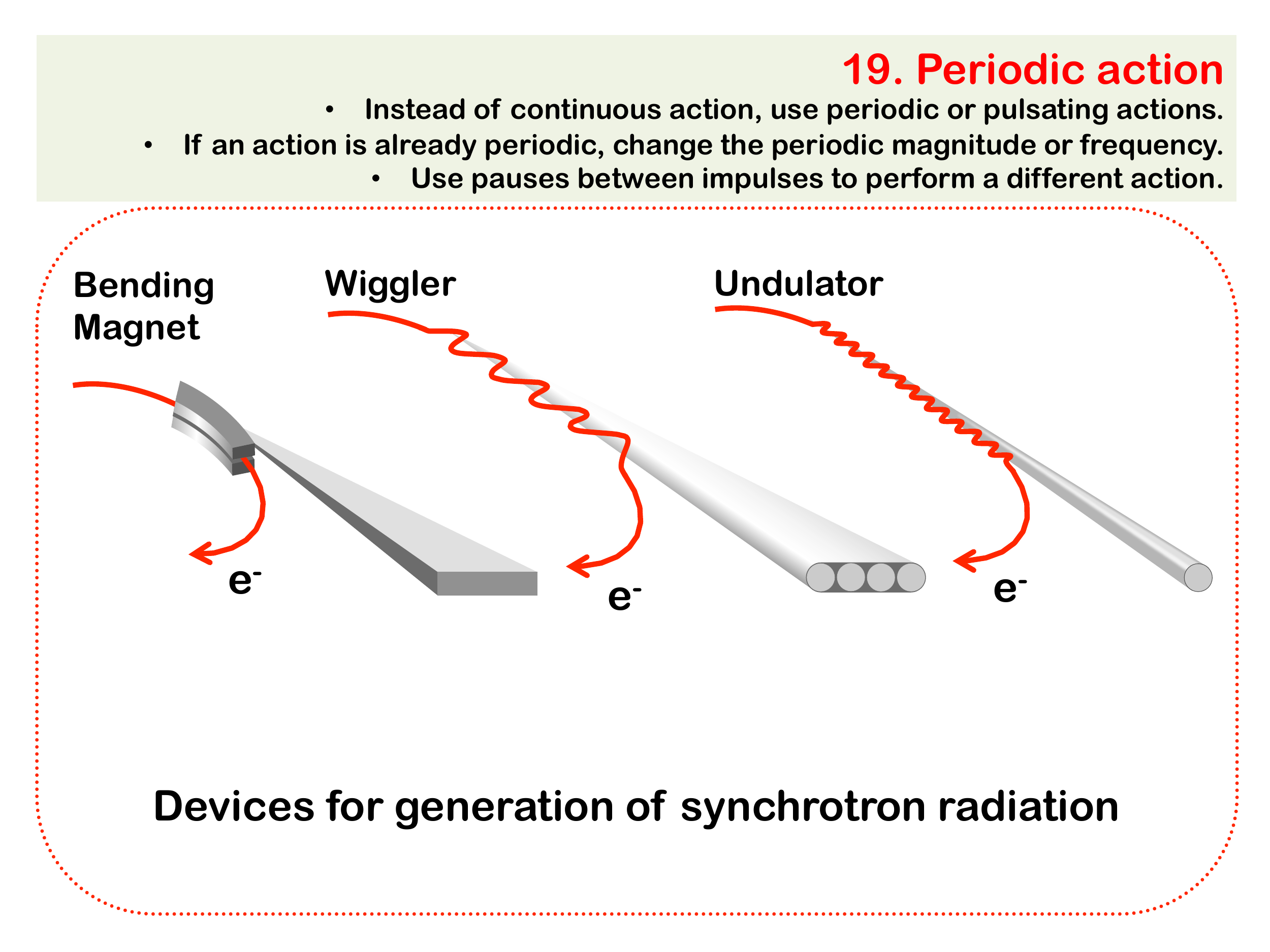}
\caption{Inventive principle ``Periodic action''.}
\label{lab19}
\end{figure}

We illustrate this principle via consideration of devices for generation of synchrotron radiation. This radiation is generated when relativistic charged particles move on a curved trajectory, loosing parts of its electromagnetic field. The simplest way to generate such radiation is to pass particles via a bending magnet, as shown on the left side of the picture. However, much better characteristic of radiation (brightness, etc) can be obtained if this process is repeated -- i.e. the particles are passed through a sequence of bends of different polarity. Such arrangements of bends are called wigglers and undulators and are now widely used in synchrotron radiation light sources.

\newpage
\section{Continuity of useful action}

The inventive principle {\it continuity of useful action} may involve carrying on work continuously, making all parts of an object work at full load, all the time, eliminating all idle or intermittent actions or work.

\begin{figure}[!h]
\hspace*{-1pc}
\includegraphics[width=1.05\textwidth]{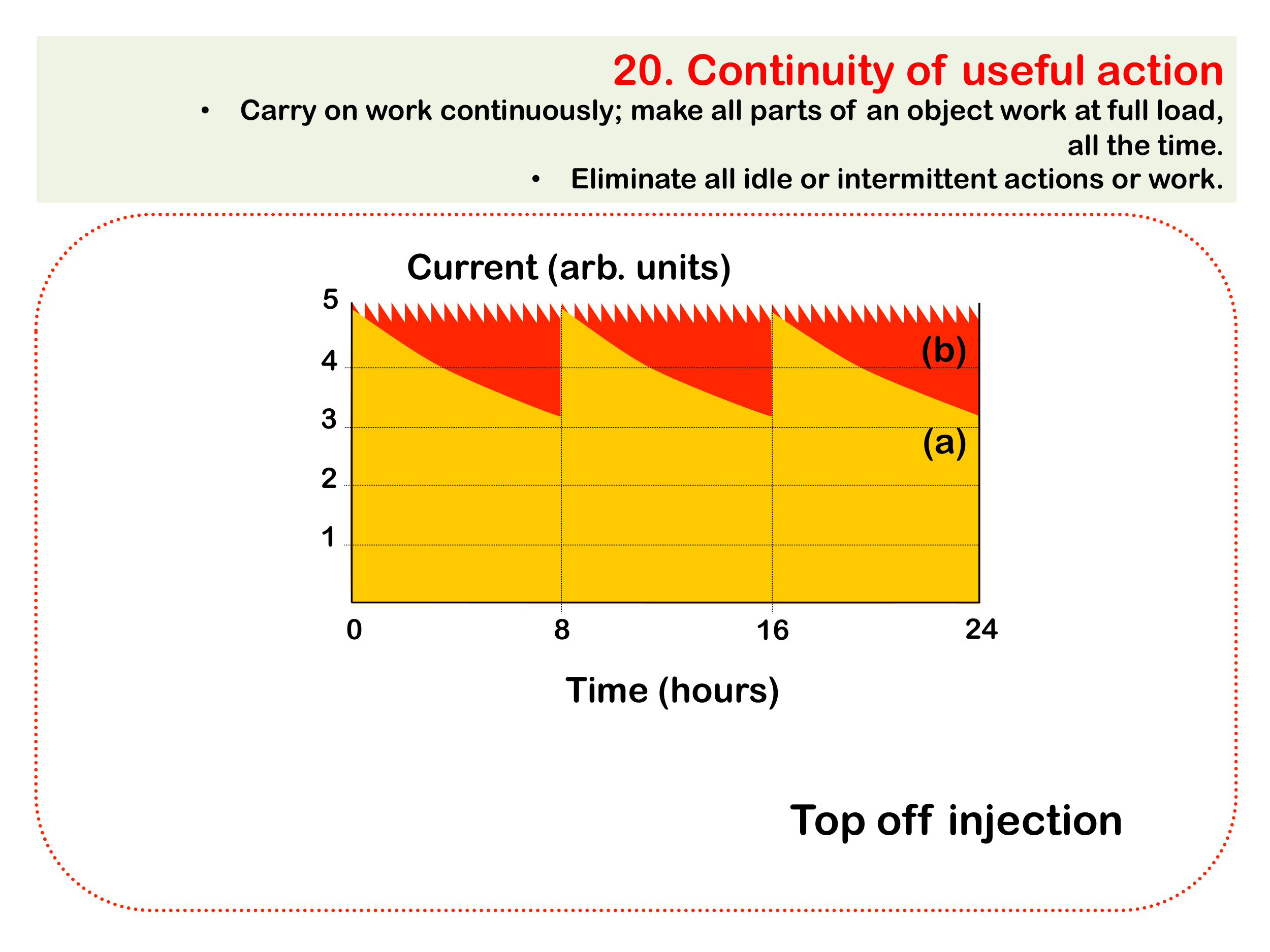}
\caption{Inventive principle ``Continuity of useful action''.}
\label{lab20}
\end{figure}

We illustrate this principle via a concept of top-off injection for synchrotron light sources. In these sources radiation is generated by circulating electron beam which can decay due to losses, and thus the circulating current as well as the intensity of generated radiation decrease with time, as shown on the left side of the picture. The circulating beam is renewed seldom, when the new beam is injected.  

An alternative way to operate the light source is to arrange almost continuous injection so that the fresh portion of the beam would be injected very often, considerably increasing the efficiency of the light source and also eliminating any thermal effects associated with variations of circulating current or intensity of the emitted radiation. 

\newpage
\section{Skipping}

The inventive principle {\it skipping} may involve conducting a process, or certain stages of it (e.g. destructible, harmful or hazardous operations) at high speed.

\begin{figure}[!h]
\hspace*{-1pc}
\includegraphics[width=1.05\textwidth]{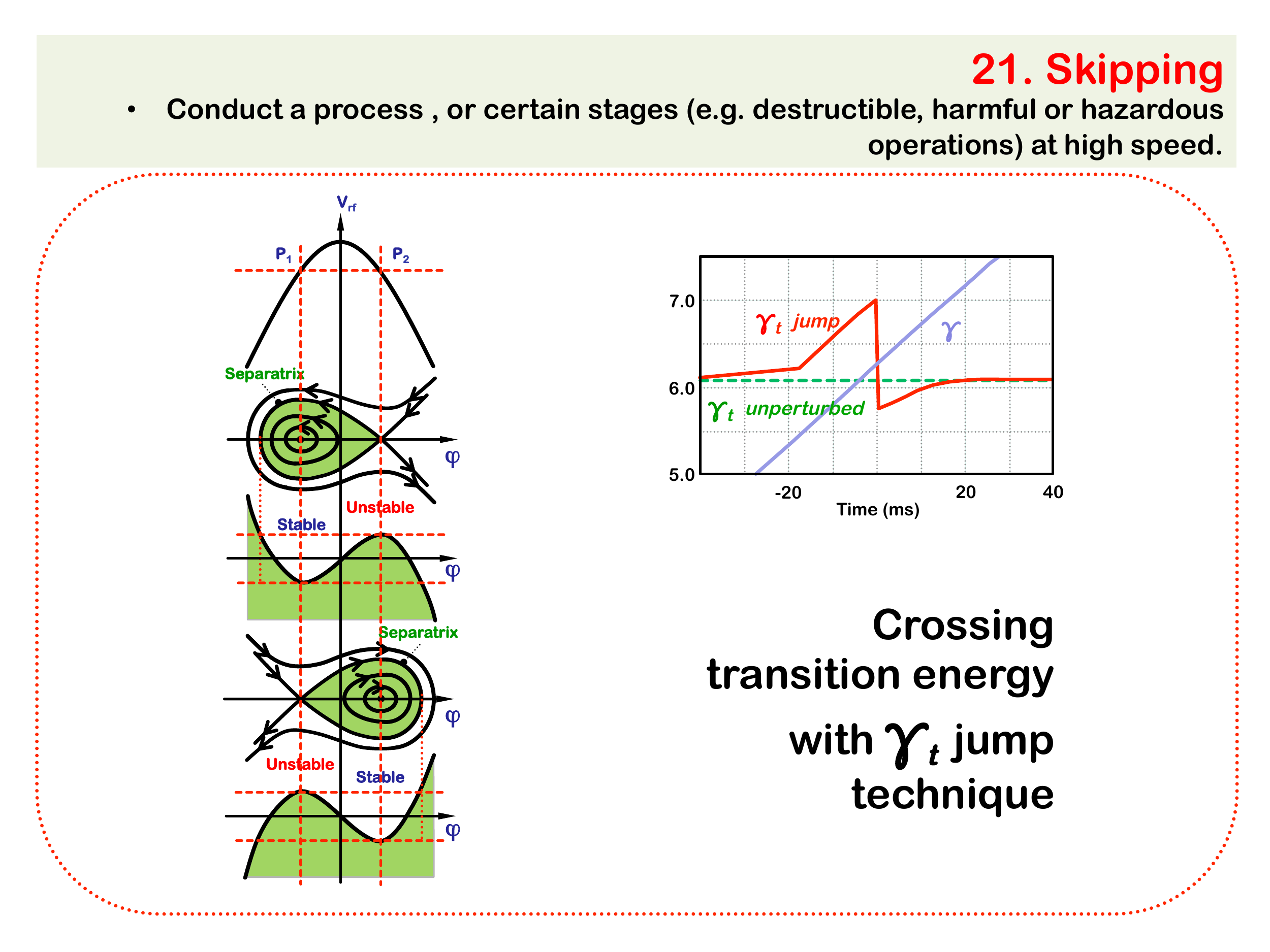}
\caption{Inventive principle ``Skipping''.}
\label{lab21}
\end{figure}

An example we selected to illustrate this principle is a gamma-jump technique used in accelerators. In proton circular accelerators in particular there is a notion of critical energy -- at this energy (the relativistic factor corresponding to this energy  is called $\gamma_t$) the stable phase (of the accelerating resonator) flips to the other side of the sine wave. Passing the critical energy during acceleration therefore requires jumping the phase of the resonator, to avoid beam getting lost. Still, some disturbance of the beam unavoidably happens, due to such transition. 

The critical energy value depends on the properties of the focusing optics of the accelerator. It is possible, therefore, to program the optics change in such a way, that transition through the critical energy will happen much faster, significantly reducing the detrimental effects on the accelerated beam.

\newpage
\section{Blessing in disguise}

The inventive principle {\it blessing in disguise} (which is also called {\it turning lemons into lemonade}) may involve using harmful factors (particularly, harmful effects of the environment or surroundings) to achieve a positive effect, eliminating the primary harmful action by adding it to another harmful action to resolve the problem, or amplifying a harmful factor to such a degree that it is no longer harmful.

\begin{figure}[!h]
\hspace*{-1pc}
\includegraphics[width=1.05\textwidth]{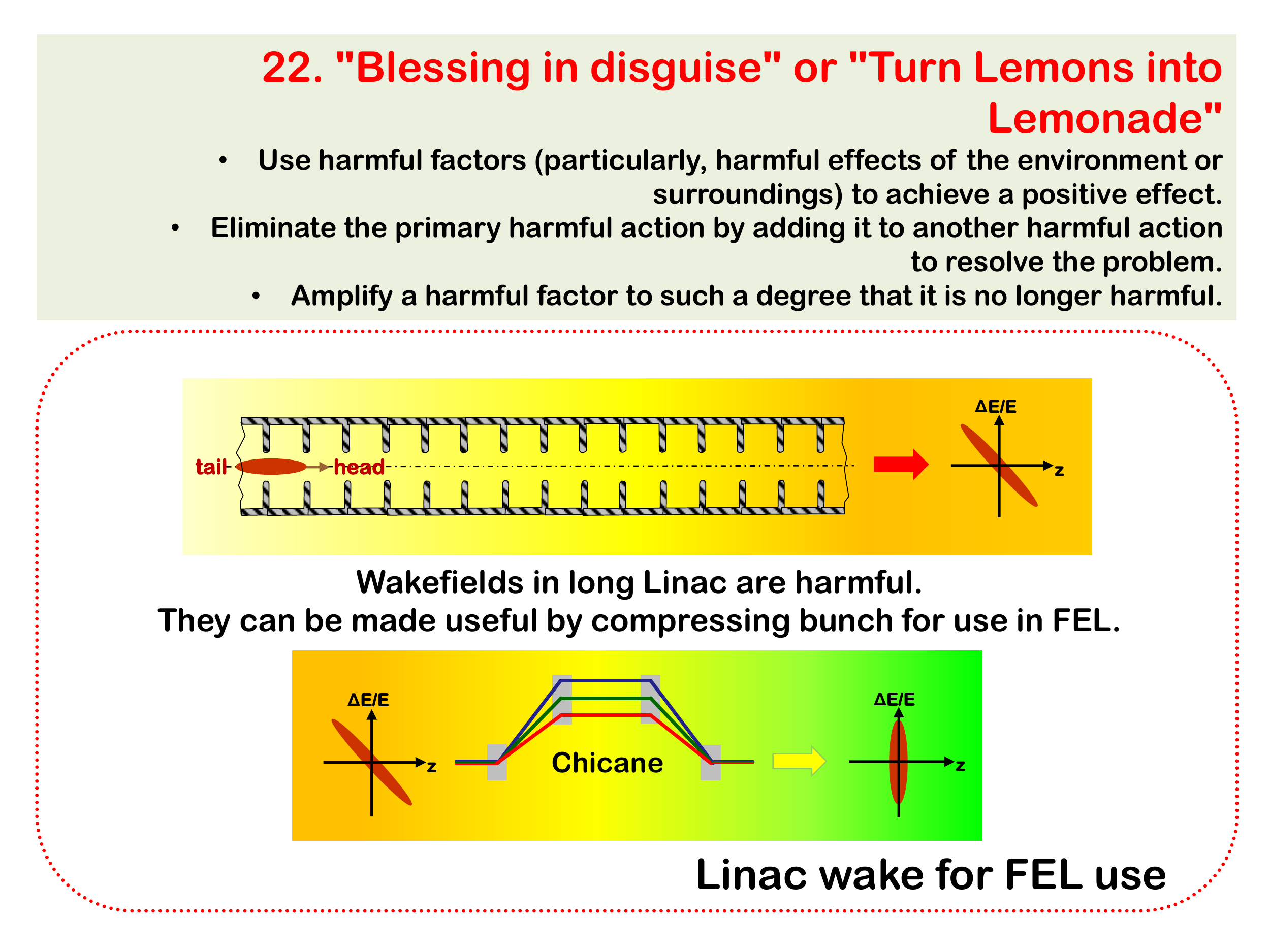}
\caption{Inventive principle ``Blessing in disguise'' or ``Turn Lemons into Lemonade''.}
\label{lab22}
\end{figure}

We illustrate this principle via consideration of wake-fields in linacs, which are normally harmful, as they can deteriorate the quality of accelerated beam, however in the case when this linac feeds a free electron laser, such wakes can turn into a useful factor, as they  can produce an energy chirp along the beam which can serve to further compress the beam longitudinally, enhancing the generated radiation.

\newpage
\section{Feedback}

The inventive principle {\it feedback} may involve introducing feedback of feed-forward links into the process (referring back, cross-checking) to improve the process or an action.

\begin{figure}[!h]
\hspace*{-1pc}
\includegraphics[width=1.05\textwidth]{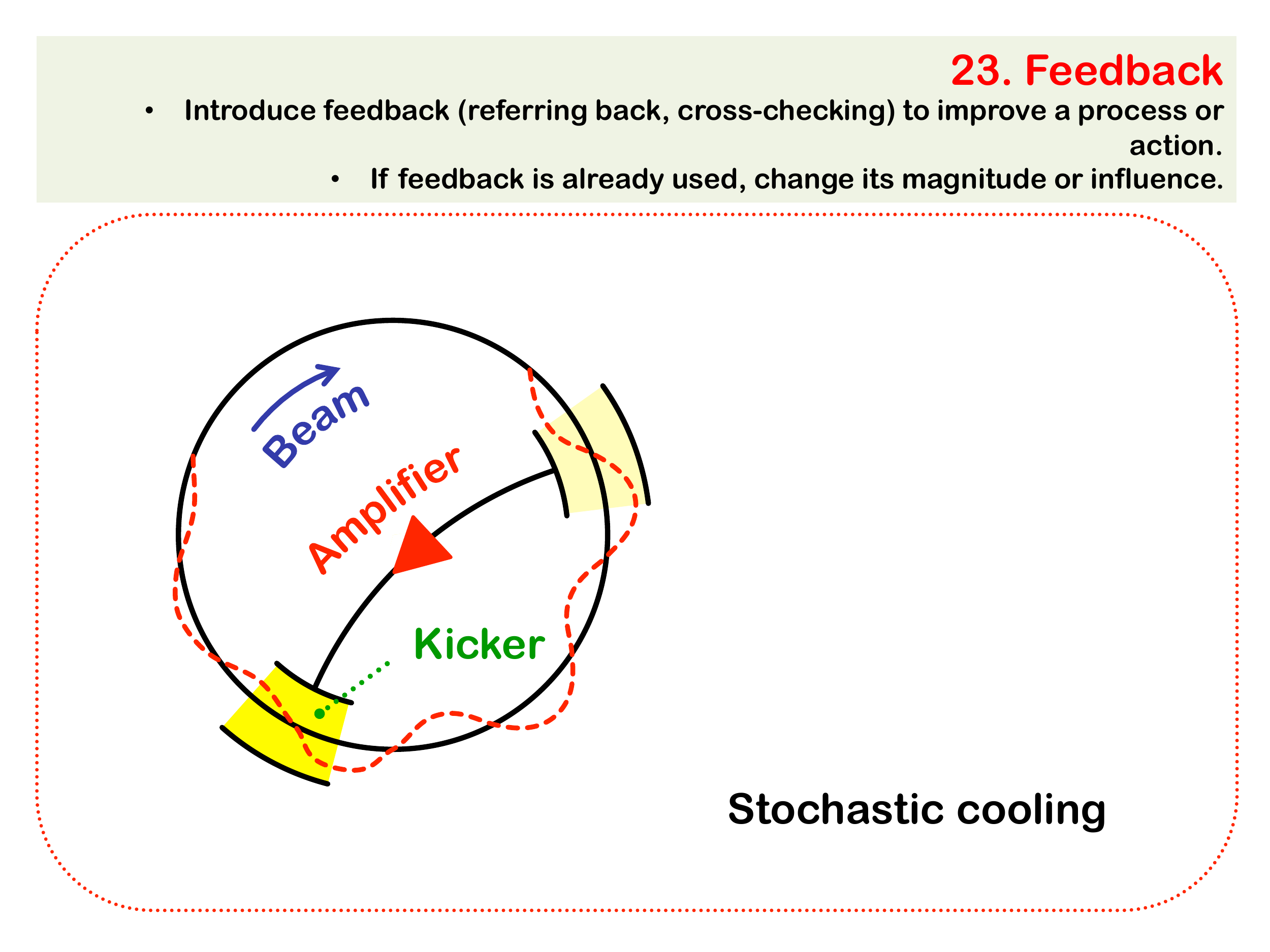}
\caption{Inventive principle ``Feedback''.}
\label{lab23}
\end{figure}

An example we selected to illustrate this principle is a concept of stochastic cooling intended for reduction of phase space volume of the circulating in an accelerator beam. This is done via first detecting oscillations of particles in one location, sending the amplified signal which carries information about these oscillation along a shorter path than the particle takes to travel, and acting on the same particle with a kick that would decrease its oscillation.

\newpage
\section{Intermediary}

The inventive principle {\it intermediary} may involve using an intermediary carrier object or intermediary process, or merging one object temporarily with another (which can be easily removed). 

\begin{figure}[!h]
\hspace*{-1pc}
\includegraphics[width=1.05\textwidth]{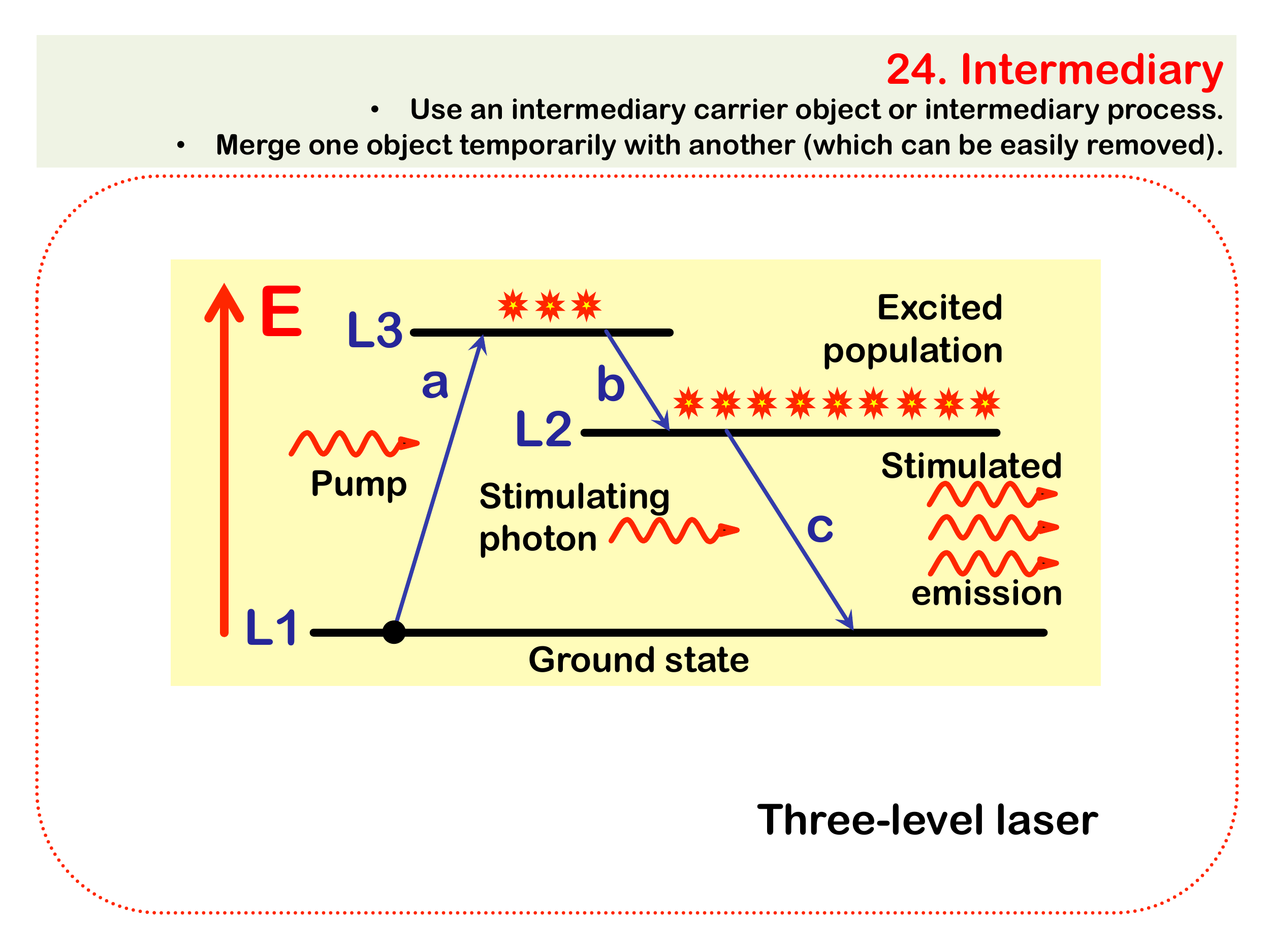}
\caption{Inventive principle ``Intermediary''.}
\label{lab24}
\end{figure}

An example we selected to illustrate this principle is a principle of operation of three-level laser. In such laser the atoms of the active media at first are located in ground state, and then, due to excitation by pump, get to the level L3, which has quite short life-time, so the atoms quickly revert, via non-radiative decay, to level L2, which has, in contrast, long life time. Any stimulating emission then would result in an avalanche of stimulated coherent radiation. The level L3, as we see in this case, plays the role of an intermediary.

\newpage
\section{Self service}

The inventive principle {\it self service} may involve making an object serve itself by performing auxiliary helpful functions, or using waste resources, energy, or substances. 

\begin{figure}[!h]
\hspace*{-1pc}
\includegraphics[width=1.05\textwidth]{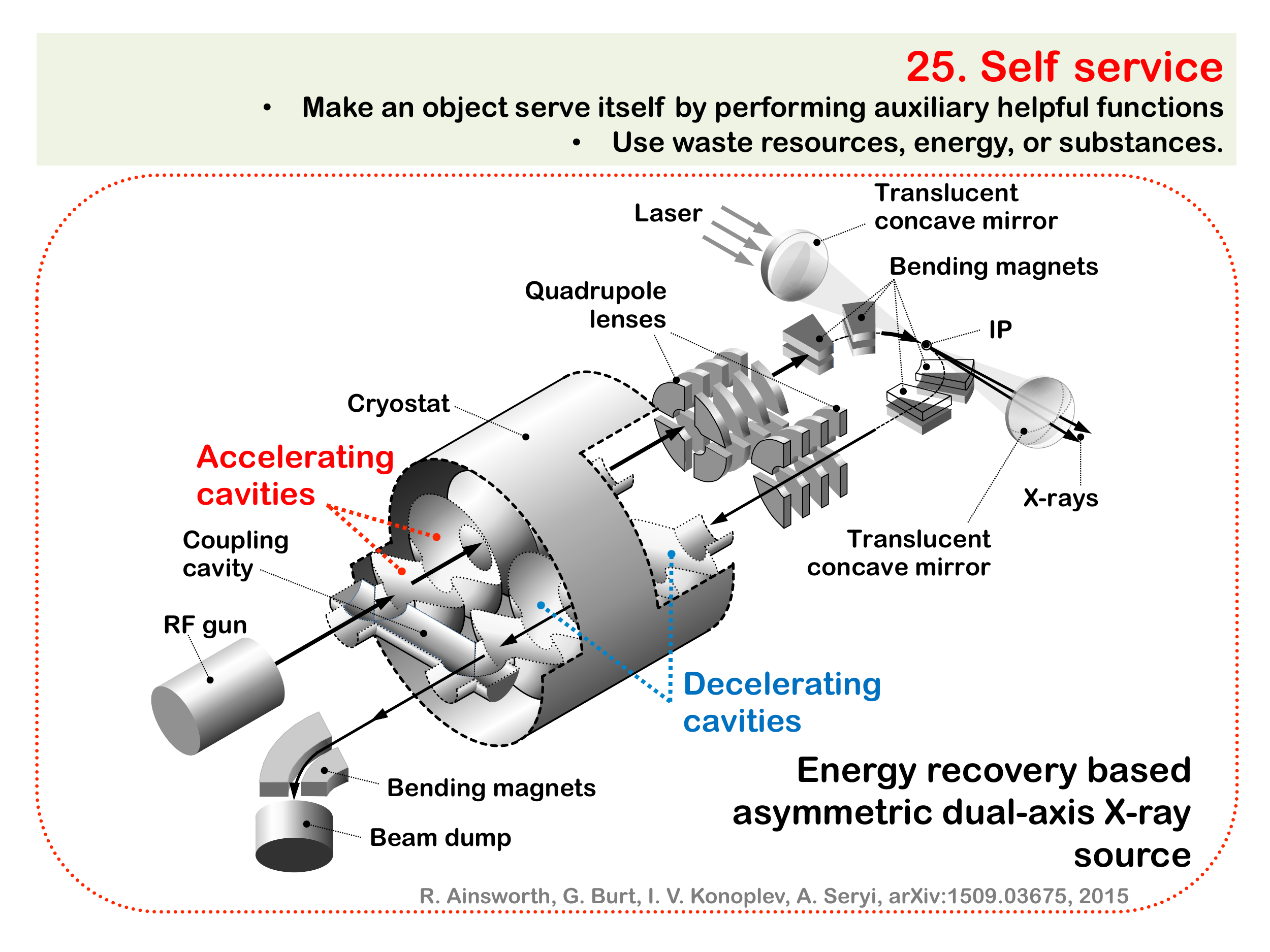}
\caption{Inventive principle ``Self service''.}
\label{lab25}
\end{figure}

An example we selected to illustrate this principle is a design of an energy recovery based asymmetric dual-axis X-ray source\footnote {R. Ainsworth, G. Burt, I. V. Konoplev, A. Seryi, {\it Asymmetric Dual Axis Energy Recovery Linac for Ultra-High Flux sources of coherent X-ray/THz radiation: Investigations Towards its Ultimate Performance}, arXiv:1509.03675, physics.acc-ph, Sep 2015.}, where the cavities are arranged in such a way as to perform an auxiliary helpful function of decelerating the beam which has already produced useful X-ray radiation, in order to recuperate its energy and thus allow more efficient operation of the system. 

\newpage
\section{Copying}

The inventive principle {\it copying} may involve using, instead of an unavailable, expensive or fragile object, its simpler and inexpensive copies, or replacing an object, or process with optical copies.

\begin{figure}[!h]
\hspace*{-1pc}
\includegraphics[width=1.05\textwidth]{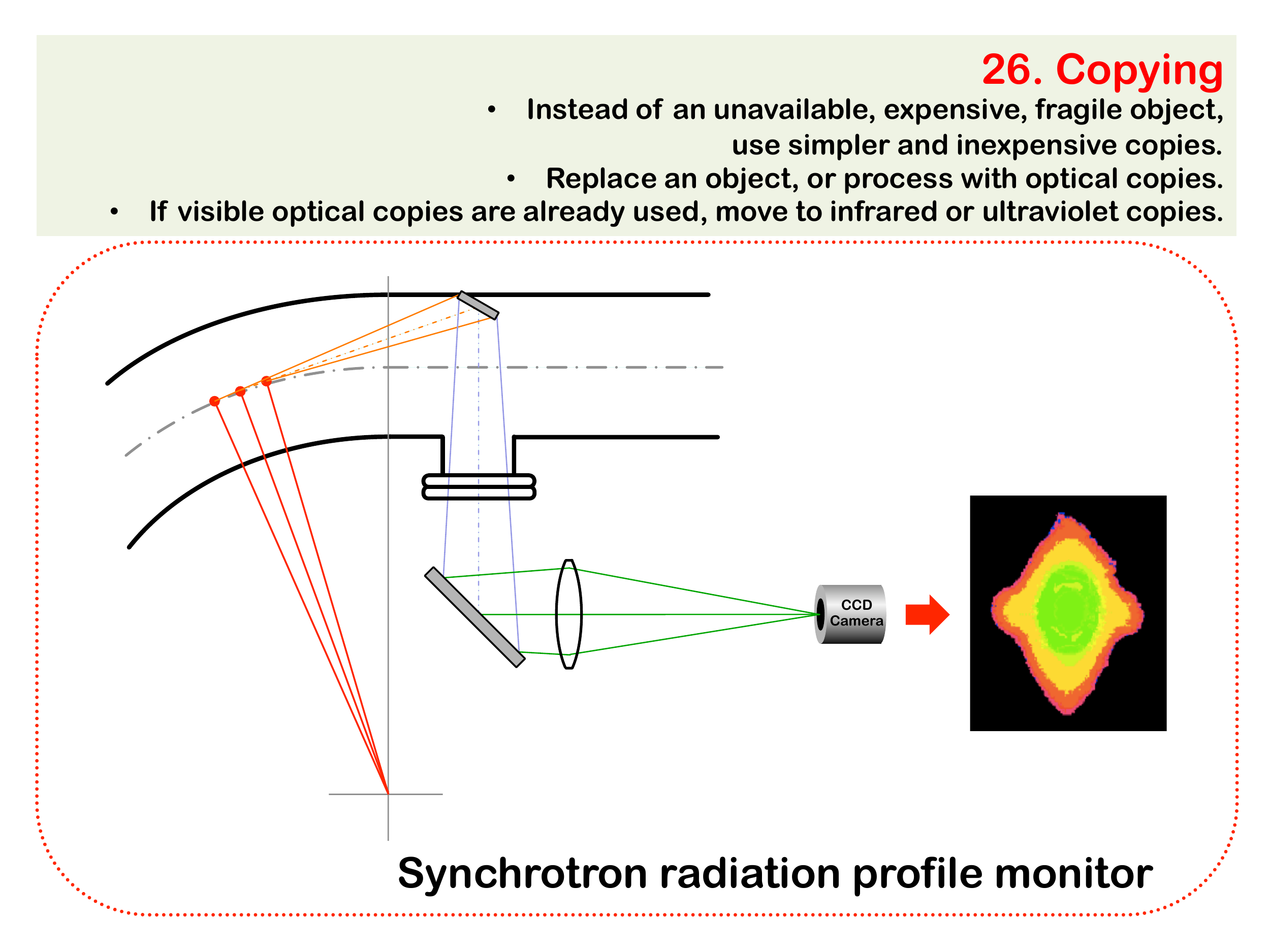}
\caption{Inventive principle ``Copying''.}
\label{lab26}
\end{figure}

An example we selected to illustrate this principle is a concept of synchrotron light beam size monitor. The beam of charged particles, circulating in an accelerator, emits synchrotron radiation, when it is passing bending magnets. Beam size monitors such as wires crossing the beams are invasive, and should be avoided. Synchrotron radiation however allows to create non-destructive monitors, when this radiation is directed with mirrors to detector and its profile analyzed. Therefore, instead of fragile beam which could be analyzed with a crossing wire, its optical copy is analyzed in this case to determine its beam size. 

\newpage
\section{Cheap short-living objects}

The inventive principle {\it cheap short-living objects} may involve replacing an expensive object with a multiple of inexpensive objects, comprising certain qualities (such as service life, for instance). 

\begin{figure}[!h]
\hspace*{-1pc}
\includegraphics[width=1.05\textwidth]{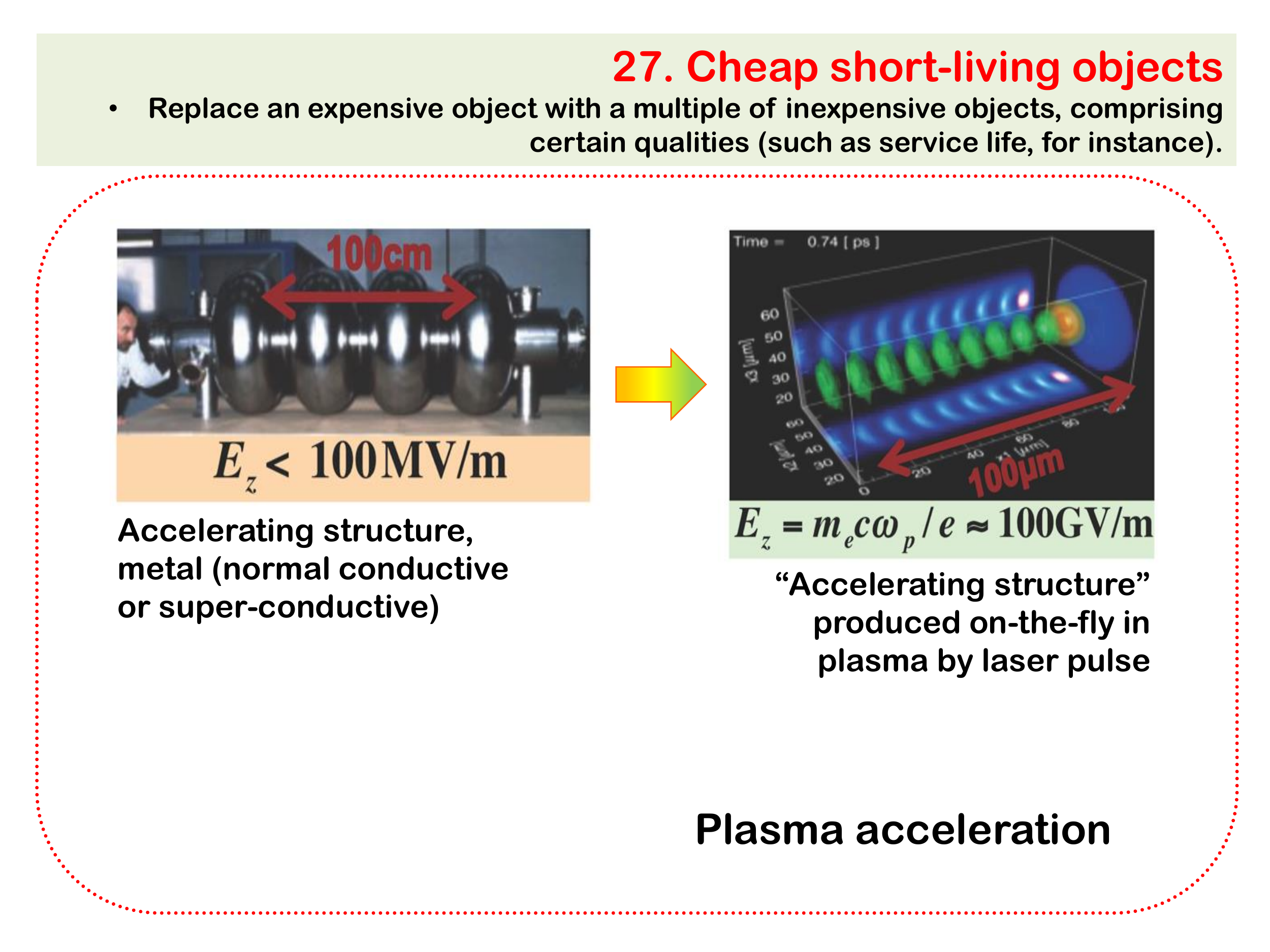}
\caption{Inventive principle ``Cheap short-living objects''.}
\label{lab27}
\end{figure}

An example we selected to illustrate this principle is a concept of laser plasma acceleration. In standard acceleration the accelerating wave is excited inside of metal resonators. The accelerating gradient is therefore limited by the properties of materials and typically cannot exceed 100~MV/m. However, a short and powerful laser pulse can excite a wave in plasma, where accelerating gradient can be thousand times higher. This wave in plasma can be used to accelerate particles. The plasma and wave in it therefore serve in this case as a cheap and short-lived object.

\newpage
\section{Mechanics substitution}

The inventive principle {\it mechanics substitution} may involve replacing mechanical means with electromagnetic or sensory (acoustic, taste or smell) means.

\begin{figure}[!h]
\hspace*{-1pc}
\includegraphics[width=1.05\textwidth]{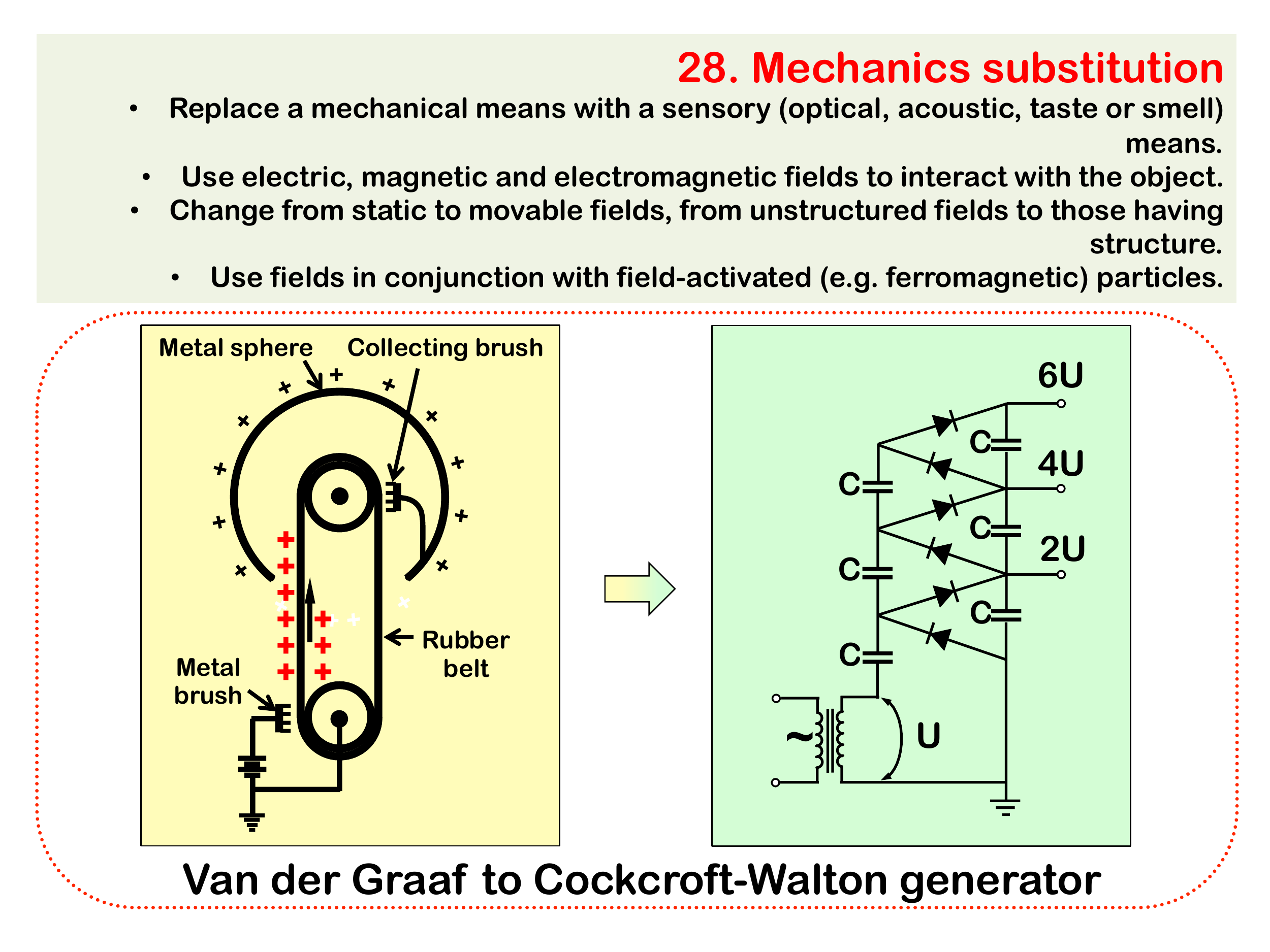}
\caption{Inventive principle ``Mechanics substitution''.}
\label{lab28}
\end{figure}

We illustrate this principle via considerations of electrostatic generators of two kinds. In the first type shown on the left part of the picture the electric charges are delivered to the metal sphere by the moving rubber belt. The charged are deposited on the belt by a system of sharp needles. This is Van der Graaf generator. In the second case a sequence of diode-based rectifiers is used to amplify the voltage to a high level, sufficient for accelerating of charged particles. This is called Cockcroft-Walton generator.

\newpage
\section{Pneumatics and hydraulics}

The inventive principle {\it pneumatics and hydraulics} may involve using gas and liquid parts of an object instead of solid parts (e.g. inflatable, filled with liquids, air cushion, hydrostatic, hydro-reactive).

\begin{figure}[!h]
\hspace*{-1pc}
\includegraphics[width=1.05\textwidth]{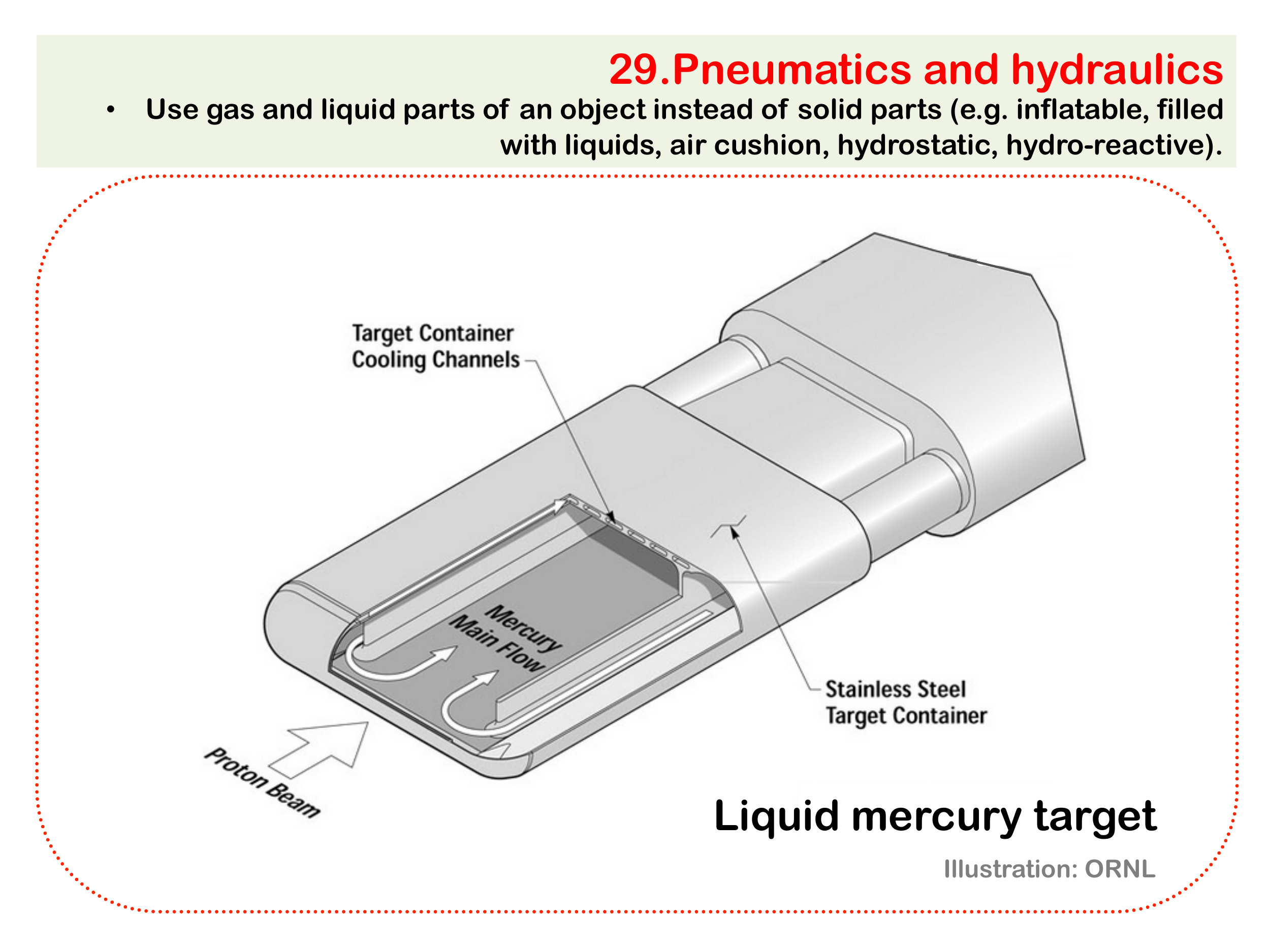}
\caption{Inventive principle ``Pneumatics and hydraulics''.}
\label{lab29}
\end{figure}

An example we selected to illustrate this principle is a design concept of a liquid beam target. Beam targets are often needed, for example, for production of positrons or neutrons from accelerated electron or proton beams. Considerable amount of heat is deposited in the targets during its interaction with the beam. Preventing destruction of the target can be done by making it rotating, or, ultimately, already ``destroyed'', i.e., in this case, liquid.

\newpage
\section{Flexible shells and thin films}

The inventive principle {\it flexible shells and thin films} may involve using flexible shells and thin films instead of three dimensional structures or 
isolating the object from the external environment using flexible shells and thin films.

\begin{figure}[!h]
\hspace*{-1pc}
\includegraphics[width=1.05\textwidth]{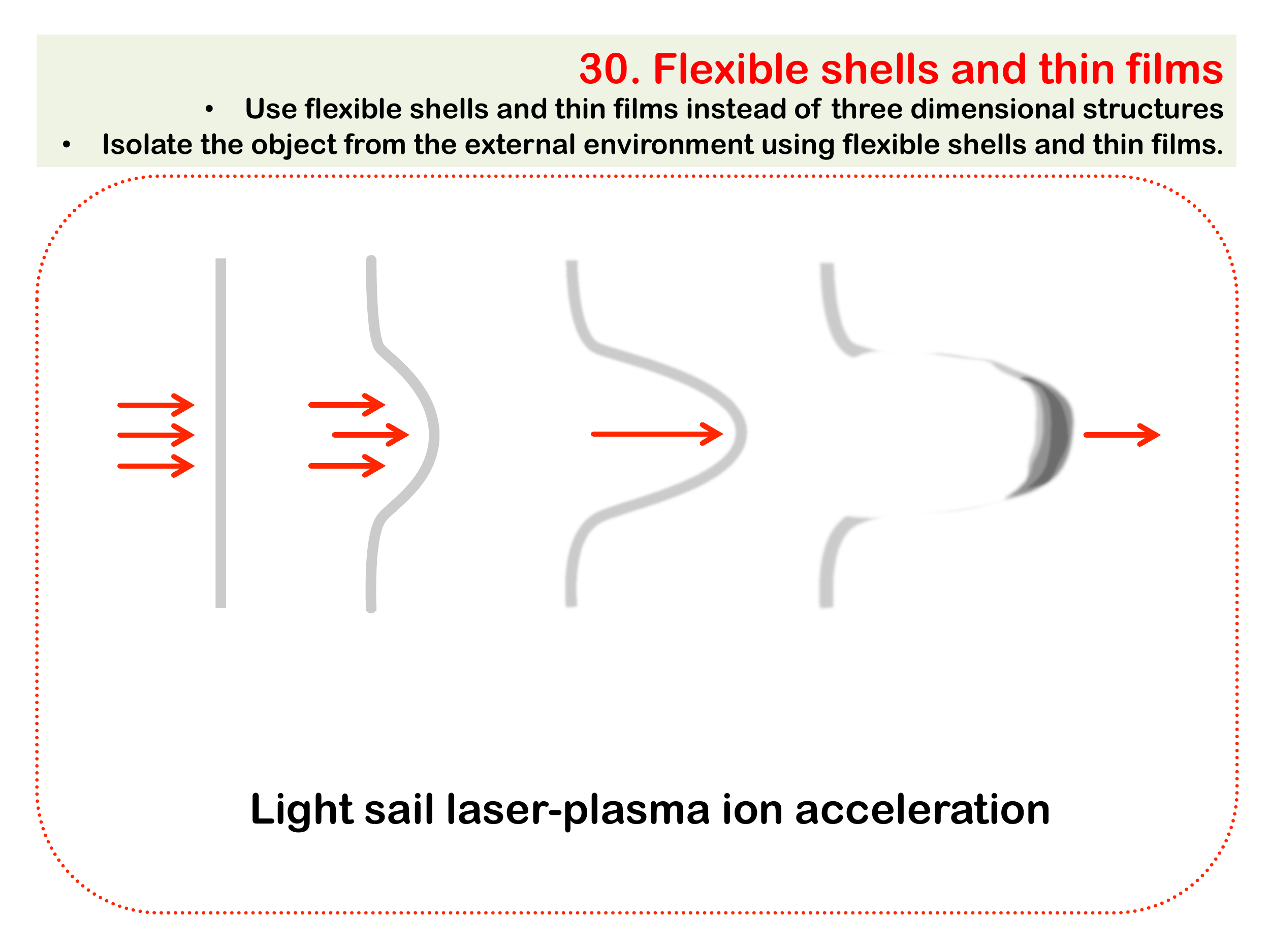}
\caption{Inventive principle ``Flexible shells and thin films''.}
\label{lab30}
\end{figure}

An example we selected to illustrate this principle is a concept of light sail laser-plasma ion acceleration. Laser plasma acceleration of ions is usually achieved via interaction of powerful short laser pulse with solid target. This method, however, does not produce a nice mono-energetic accelerated beam. From the other side, shining the laser onto a very thin film creates radiation pressure driven acceleration of the entire film, and correspondingly the ion beam accelerated in such a case will have much better properties.

\newpage
\section{Porous materials}

The inventive principle {\it porous materials} may involve making an object porous or adding porous elements (inserts, coatings, etc.), and if an object is already porous, using the pores to introduce a useful substance or function.

\begin{figure}[!h]
\hspace*{-1pc}
\includegraphics[width=1.05\textwidth]{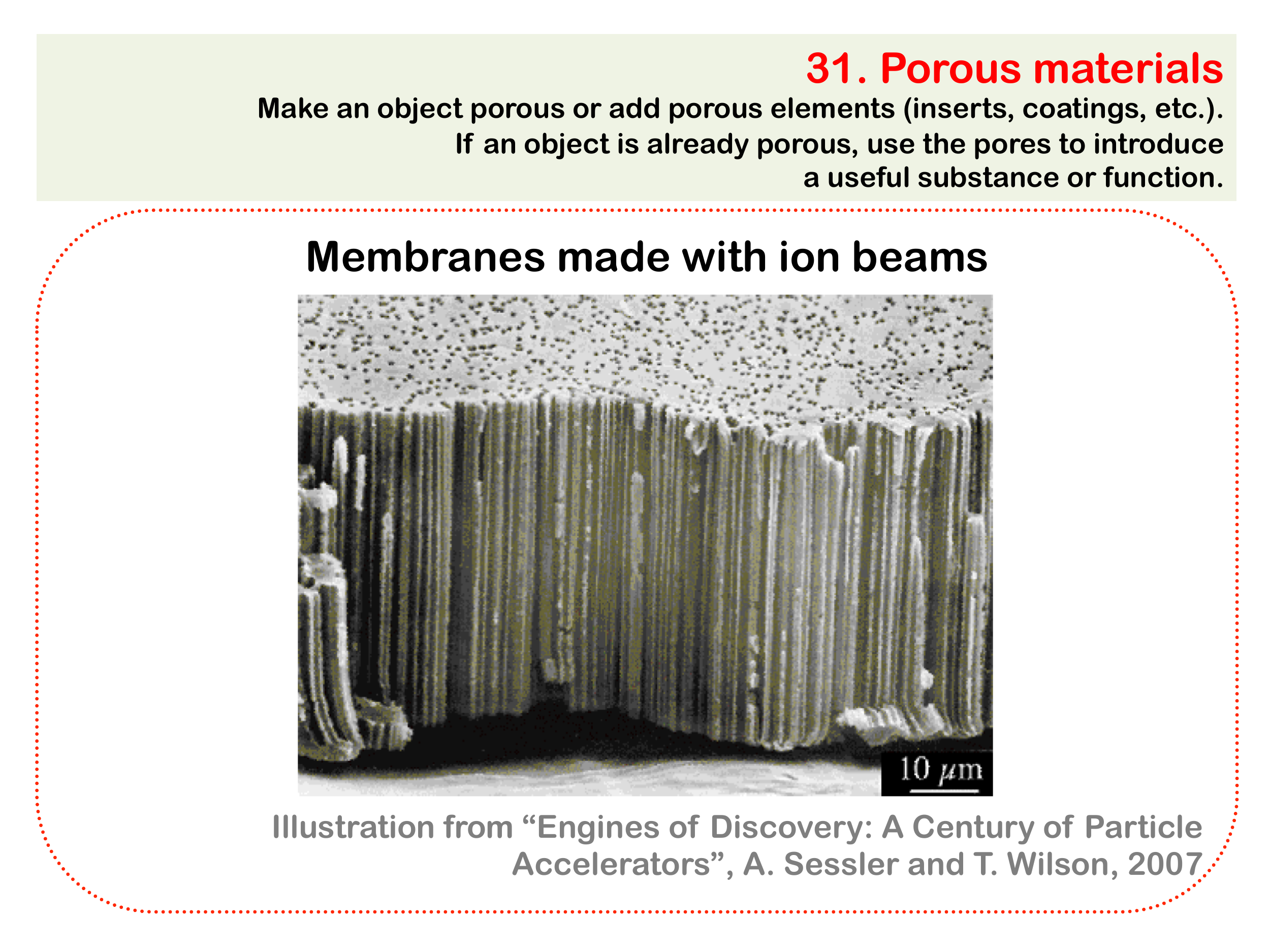}
\caption{Inventive principle ``Porous materials''.}
\label{lab31}
\end{figure}

An example we selected to illustrate this principle is a concept of creating porous membranes  using accelerated ion beams\footnote {A. Sessler and T. Wilson, {\it Engines of Discovery: A Century of Particle Accelerators}, 2007.}. In this case, ion beams accelerated typically in cyclotrons are directed into thin films, leaving ionized traces inside. Further chemical processing creates pores in such membranes, which can be then used for various filters or other applications. 

\newpage
\section{Color changes}

The inventive principle {\it color changes} may involve  changing the color of an object or its external environment, changing the transparency of an object or its external environment, changing the emissivity properties of an object subject to radiant heating, etc.

\begin{figure}[!h]
\hspace*{-1pc}
\includegraphics[width=1.05\textwidth]{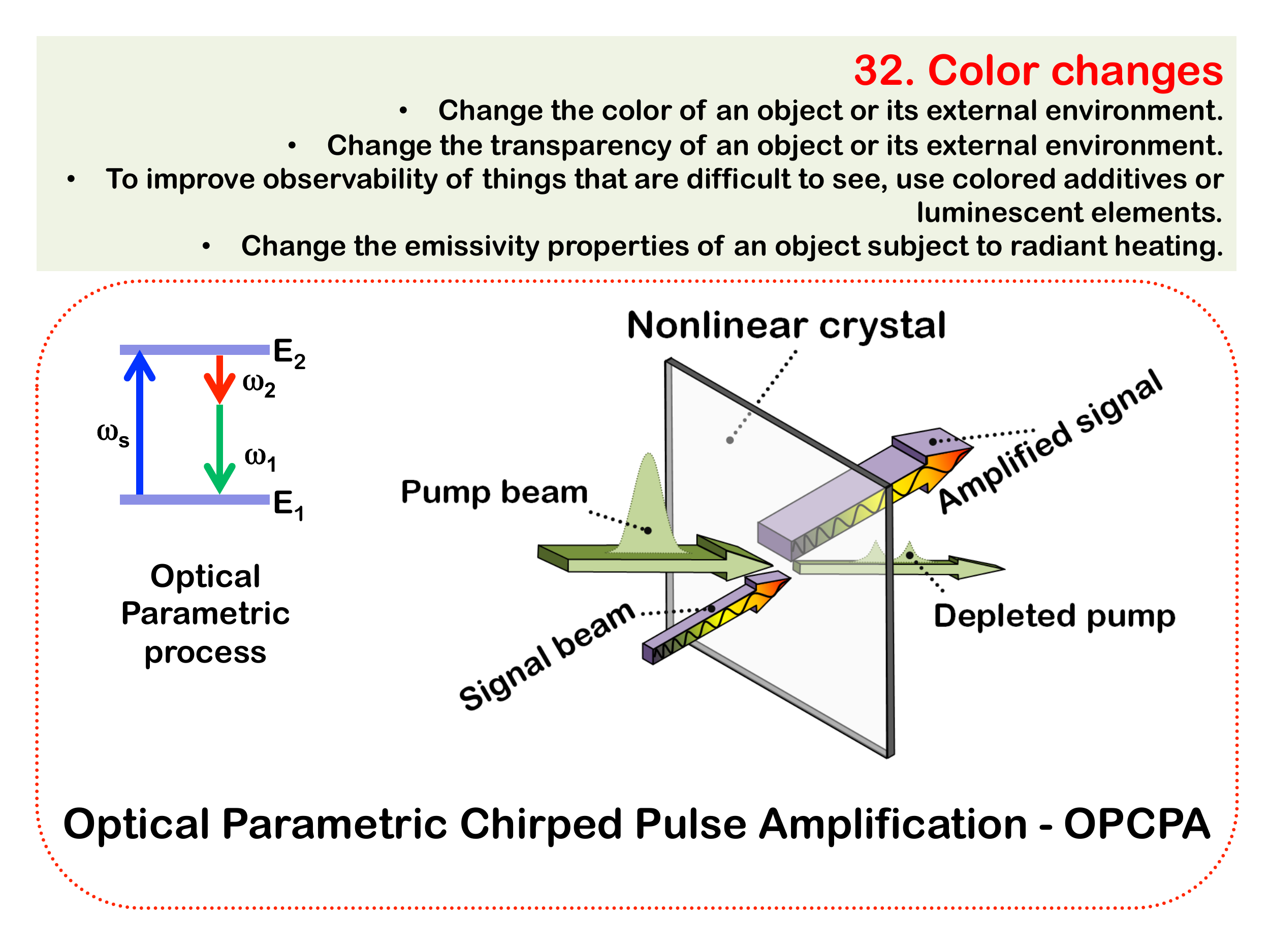}
\caption{Inventive principle ``Color changes''.}
\label{lab32}
\end{figure}

An example we selected to illustrate this principle is a principle of OPCPA  -- Optical Parametric Chirped Pulse Amplification. In this method a nonlinear crystal is used, which emits two different wavelengths when pumped with a single wavelength. This crystal can be used in an optical amplifier, when a pump laser beam and a signal laser beam are sent onto the crystal, and out from the crystal an amplified signal and depleted pump beams emerge. If the signal beam is chirped, the output amplified signal is also chirped. Change of the color by the nonlinear crystal via optical parametric process is an illustration of the color changes inventive principle.

\newpage
\section{Homogeneity}

The inventive principle {\it  homogeneity} or expressing it in Latin {\it similia similibus curantur} may involve making objects interacting with a given object of the same material (or material with identical properties).

\begin{figure}[!h]
\hspace*{-1pc}
\includegraphics[width=1.05\textwidth]{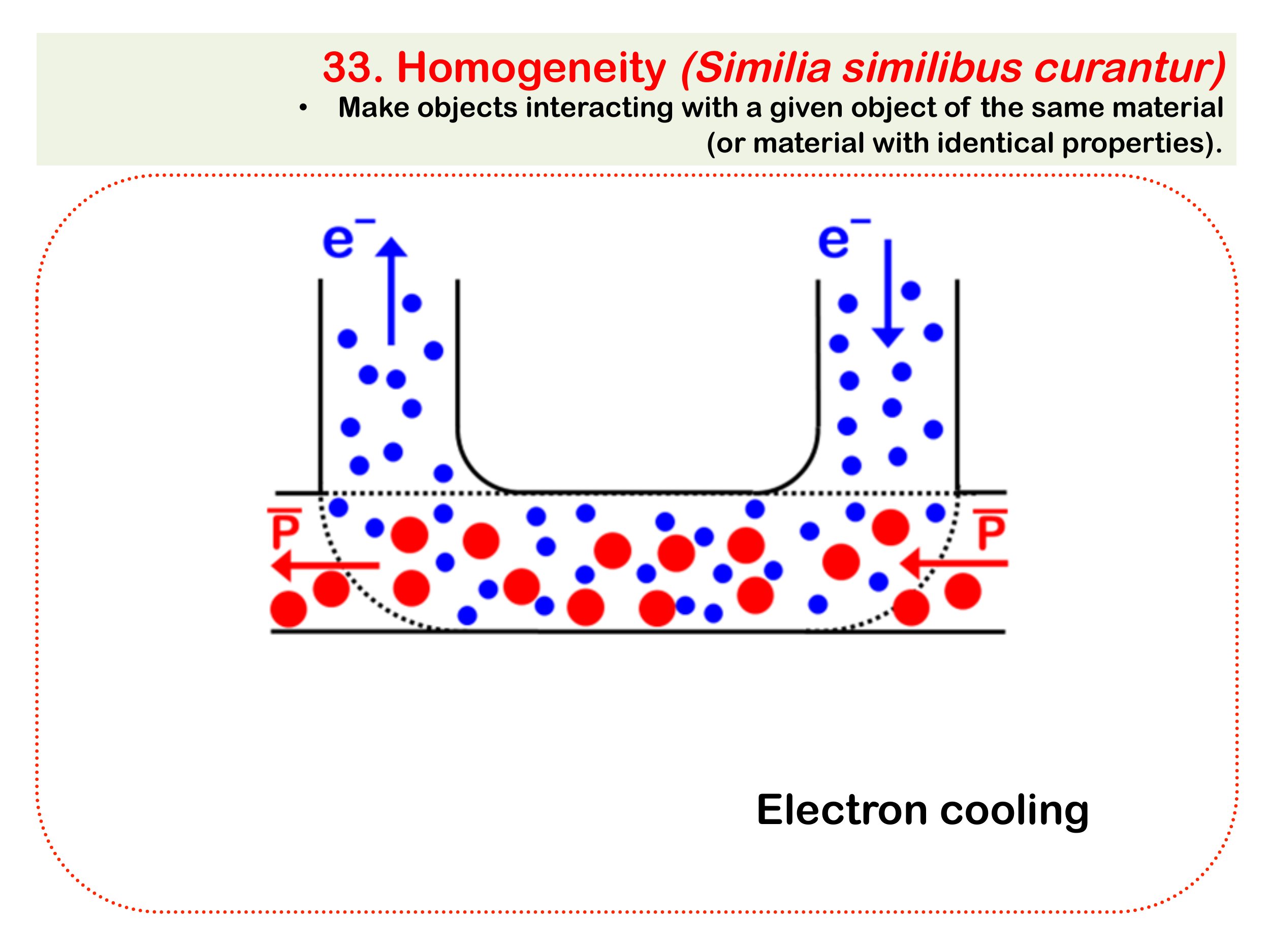}
\caption{Inventive principle ``Homogeneity''.}
\label{lab33}
\end{figure}

An example we selected to illustrate this principle is a concept of electron cooling. This cooling method is aimed at decrease of the phase space volume of charged particle beam, for example the beam of anti-protons. The cooling is done by overlapping the anti-proton beam with the beam of electrons going in the same direction and with the same velocity. Hot anti-protons, colliding with colder electrons, will transfer their energy to electrons, and after many passages through the electron beam will cool down. So, here we cure similar with similar -- cool charged particles with other type of charged particles.

\newpage
\section{Discarding and recovering}

The inventive principle {\it discarding and recovering} may involve making portions of an object that have fulfilled their functions go away (discard by dissolving, evaporating, etc.) or modifying these directly during operation, or, conversely, restoring consumable parts of an object directly in operation.

\begin{figure}[!h]
\hspace*{-1pc}
\includegraphics[width=1.05\textwidth]{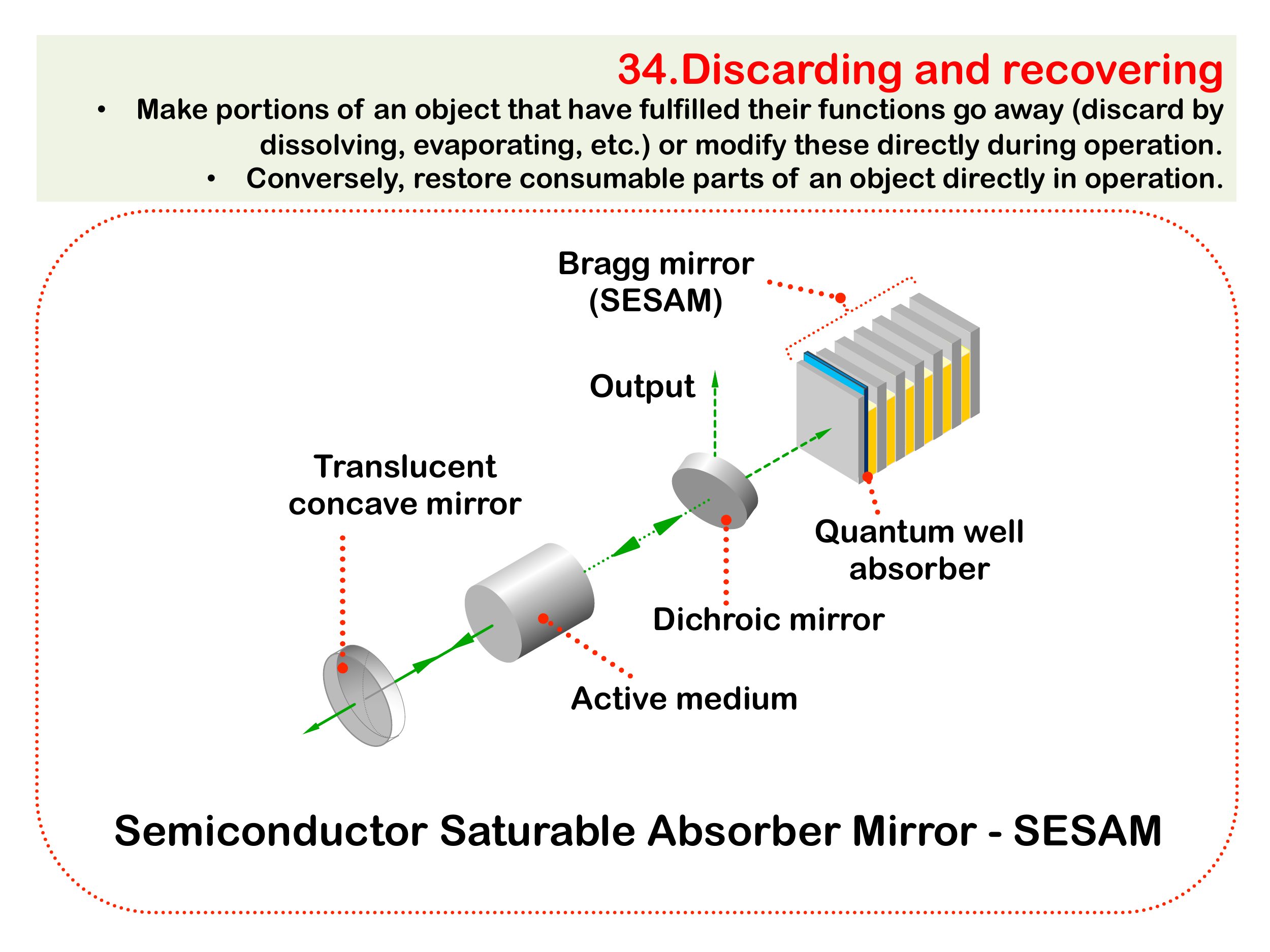}
\caption{Inventive principle ``Discarding and recovering''.}
\label{lab34}
\end{figure}

An example we selected to illustrate this principle is a Semiconductor Saturable Absorber Mirror -- so called SESAM. Such device is used in the system of short laser pulse generator and plays the role of a mirror, which, when the stored in the laser cavity light reaches certain intensity, ``disappears'' due to saturation effects, thus releasing all stored laser light out from the cavity in the form of a short intense laser pulse. The SESAM is thus a mirror that is discarded at a proper moment.

\newpage
\section{Parameter changes}

The inventive principle {\it parameter changes} may involve changing an object's physical state (e.g. to a gas, liquid, or solid), changing the concentration or consistency, changing the degree of flexibility, changing the temperature, etc. 

\begin{figure}[!h]
\hspace*{-1pc}
\includegraphics[width=1.05\textwidth]{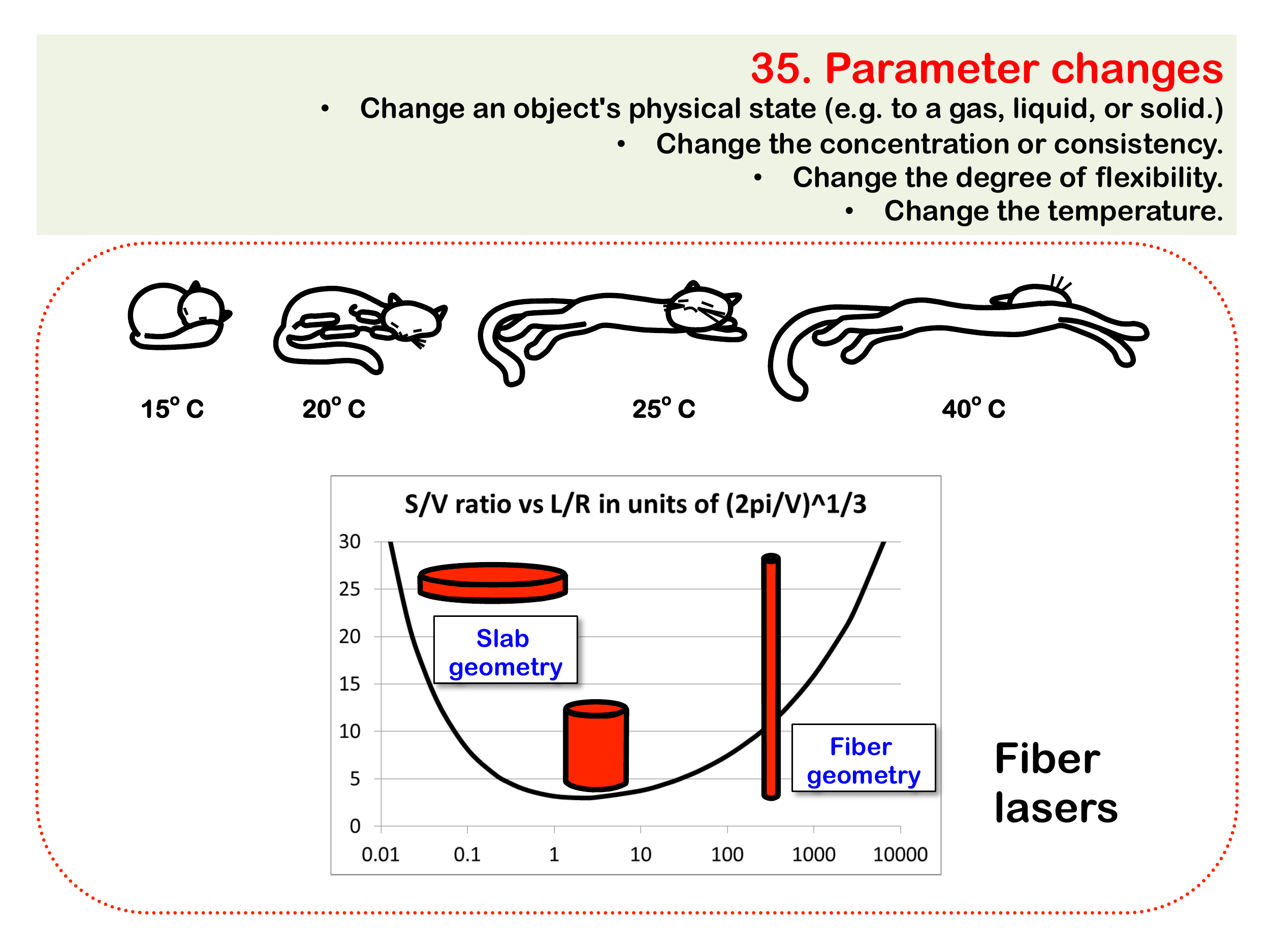}
\caption{Inventive principle ``Parameter changes''.}
\label{lab35}
\end{figure}

We illustrate this principle via consideration of variation of the ratio of the surface area of an object to the volume of the object. Maxwell or thermodynamic equations indicate that changing the volume to surface ratio can change characteristics of the object such as cooling rate or its electromagnetic field. We illustrate this via an example of a cat, who can change its surface area depending on the environmental temperature, to control its cooling rate, or via an example of fiber lasers, which, in comparison with lasers with standard geometry of active media, have much higher surface area, therefore better cooling, and correspondingly can have higher repetition rate and higher efficiency.

\newpage
\section{Phase transitions}

The inventive principle {\it phase transitions} may involve using phenomena occurring during phase transitions (e.g. volume changes, loss or absorption of heat, etc.).

\begin{figure}[!h]
\hspace*{-1pc}
\includegraphics[width=1.05\textwidth]{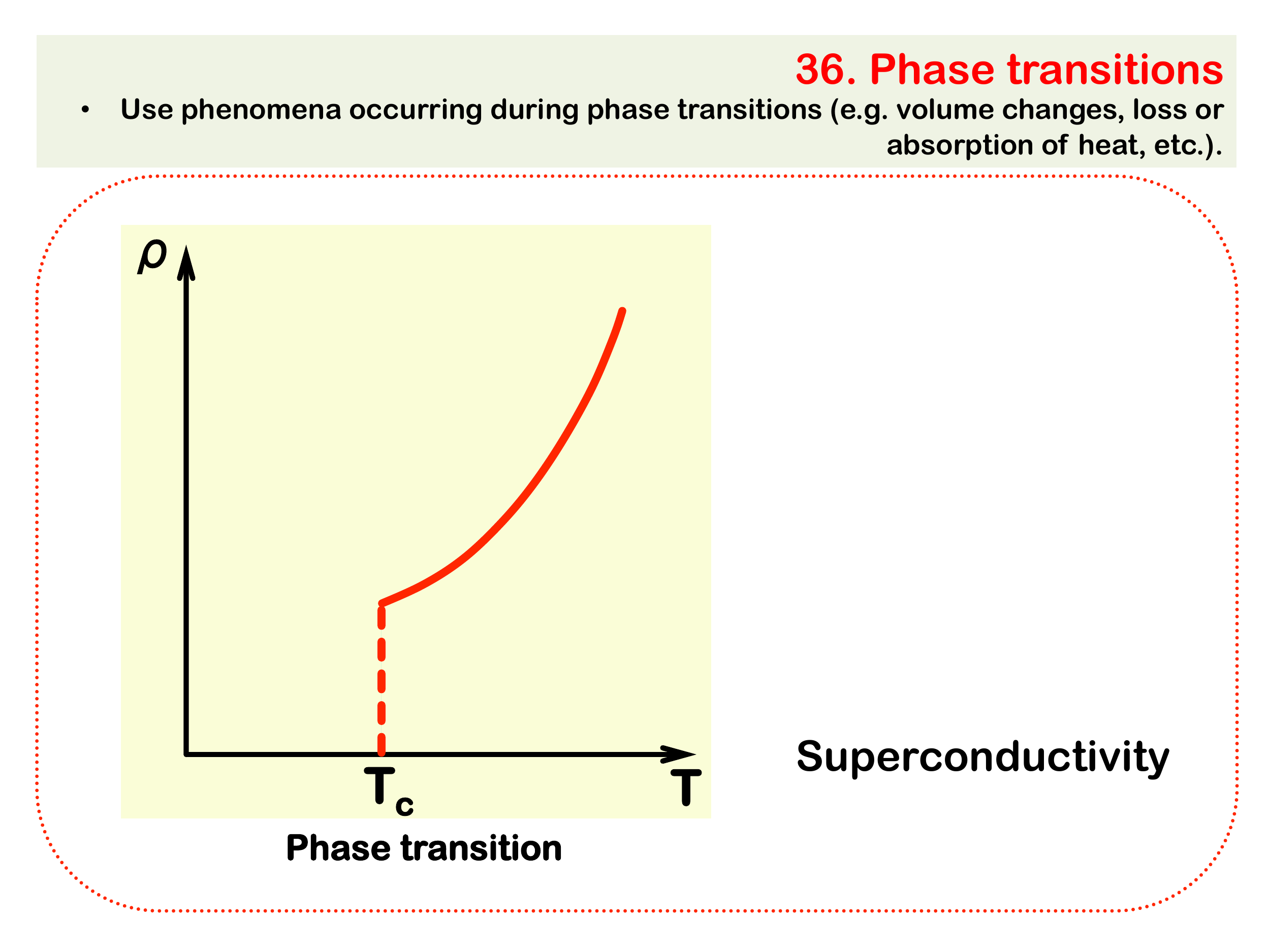}
\caption{Inventive principle ``Phase transitions''.}
\label{lab36}
\end{figure}

An example we selected to illustrate this principle is a phenomenon of superconductivity when the electrical resistance of certain materials can drop to zero when temperature decreases below the critical one. 

\newpage
\section{Thermal or electrical expansion \\or property change}

The inventive principle {\it thermal or electrical expansion or property change} may involve using thermal or electrical expansion (or contraction) or other property change of materials, or using multiple materials with different coefficients of thermal expansion (property change).

\begin{figure}[!h]
\hspace*{-1pc}
\includegraphics[width=1.05\textwidth]{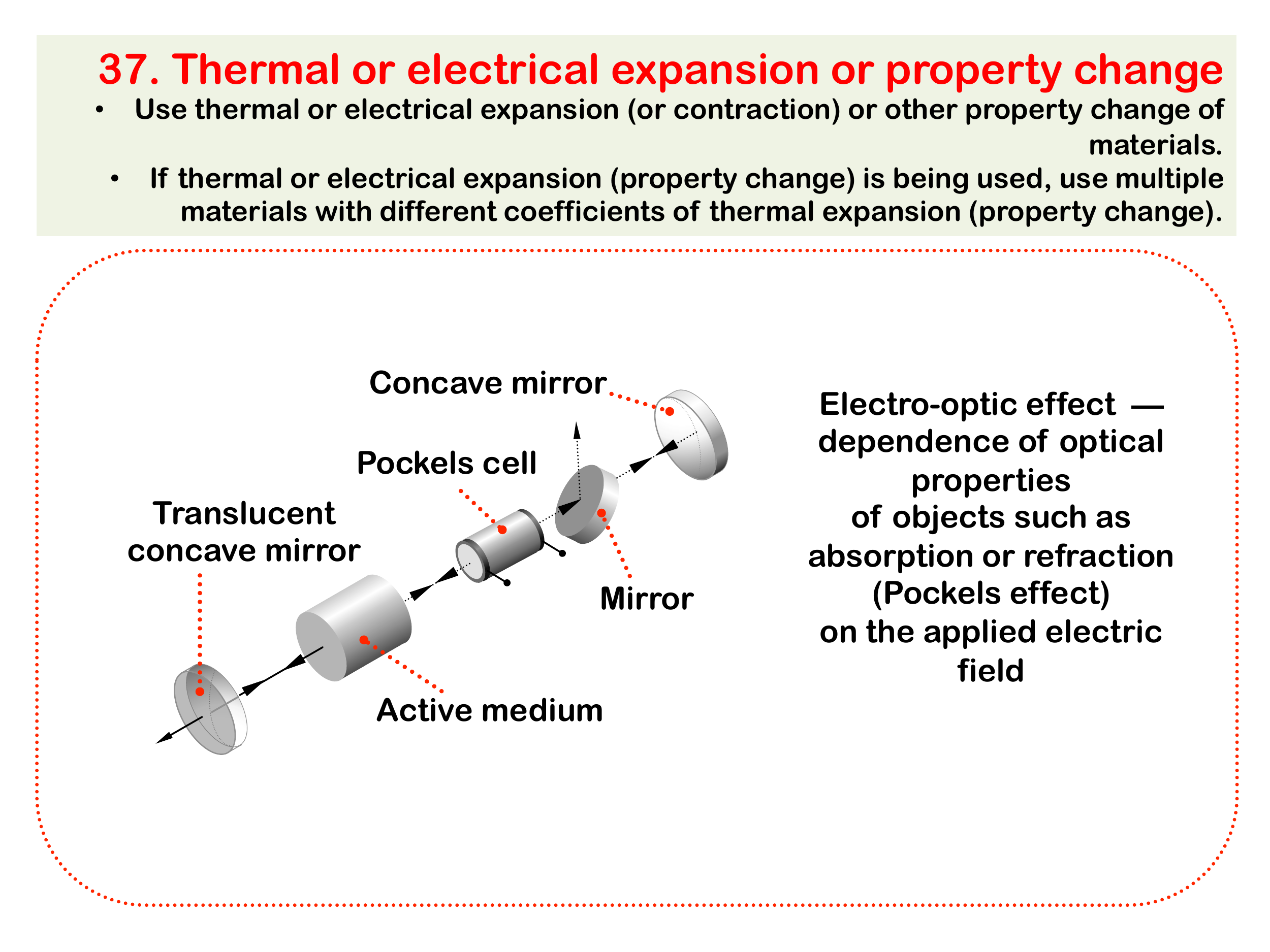}
\caption{Inventive principle ``Thermal or electrical expansion or property change''.}
\label{lab37}
\end{figure}

An example we selected to illustrate this principle is an electro-optic effect -- dependence of optical properties of objects such as absorption or refraction (Pockels effect) on the applied electric field. In the example shown on this picture this effect is used to create ultra-short laser pulses.

\newpage
\section{Strong oxidants}

The inventive principle {\it strong oxidants} may involve replacing common air with oxygen-enriched air, replacing enriched air with pure oxygen, exposing air or oxygen to ionizing radiation, using ionized oxygen or replacing ozonized (or ionized) oxygen with ozone.

\begin{figure}[!h]
\hspace*{-1pc}
\includegraphics[width=1.05\textwidth]{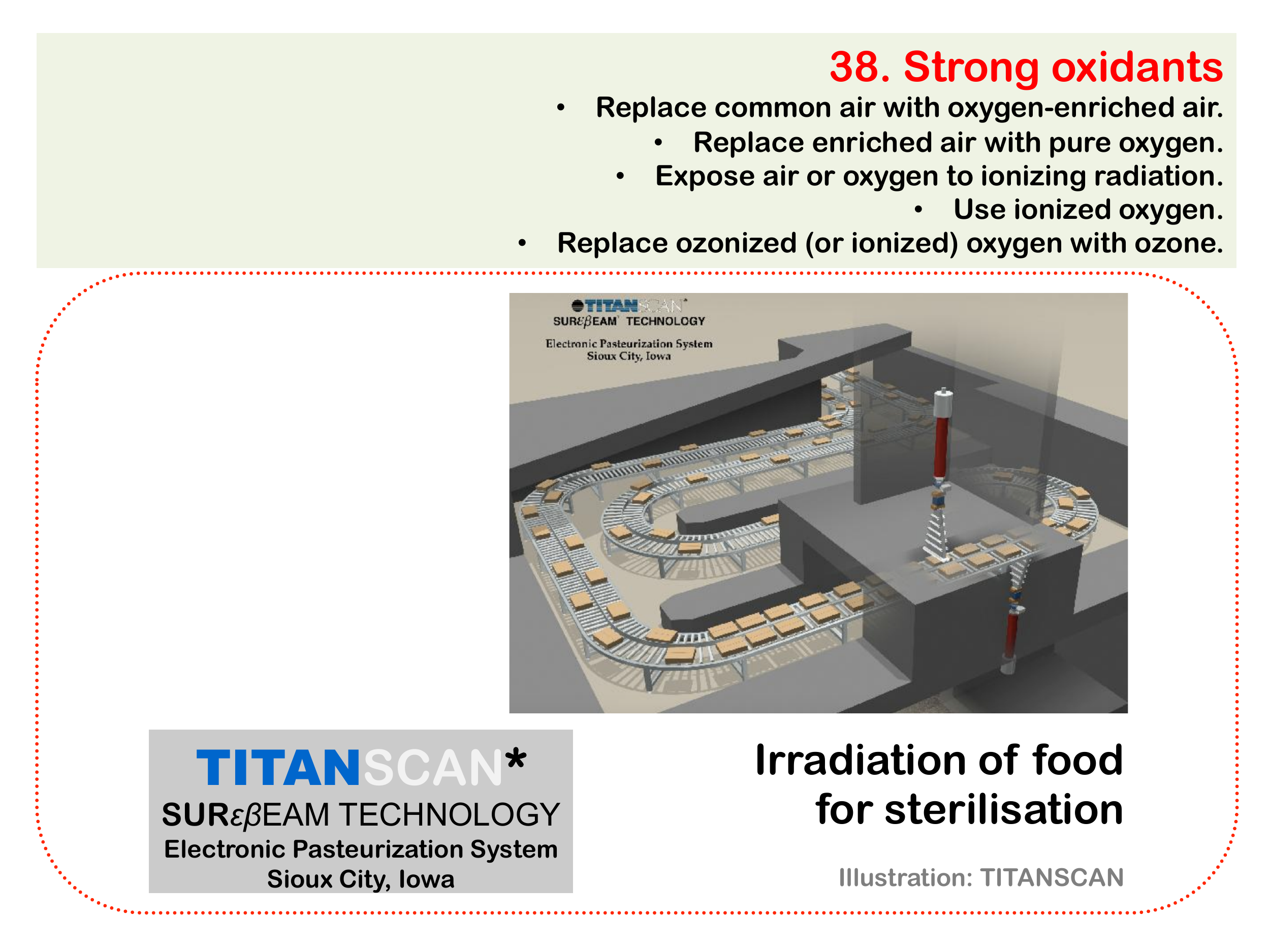}
\caption{Inventive principle ``Strong oxidants''.}
\label{lab38}
\end{figure}

An example we selected to illustrate this principle is a method for sterilization of food with low energy electron beam. Packaged food is deposited on a conveyor in a factory and passes through rastered electron beam. The picture above shows how this is done at the sterilization factory in Sioux City, Iowa. 

\newpage
\section{Inert atmosphere}

The inventive principle {\it inert atmosphere} may involve replacing a normal environment with an inert one, or adding neutral parts, or inert additives to an object.

\begin{figure}[!h]
\hspace*{-1pc}
\includegraphics[width=1.05\textwidth]{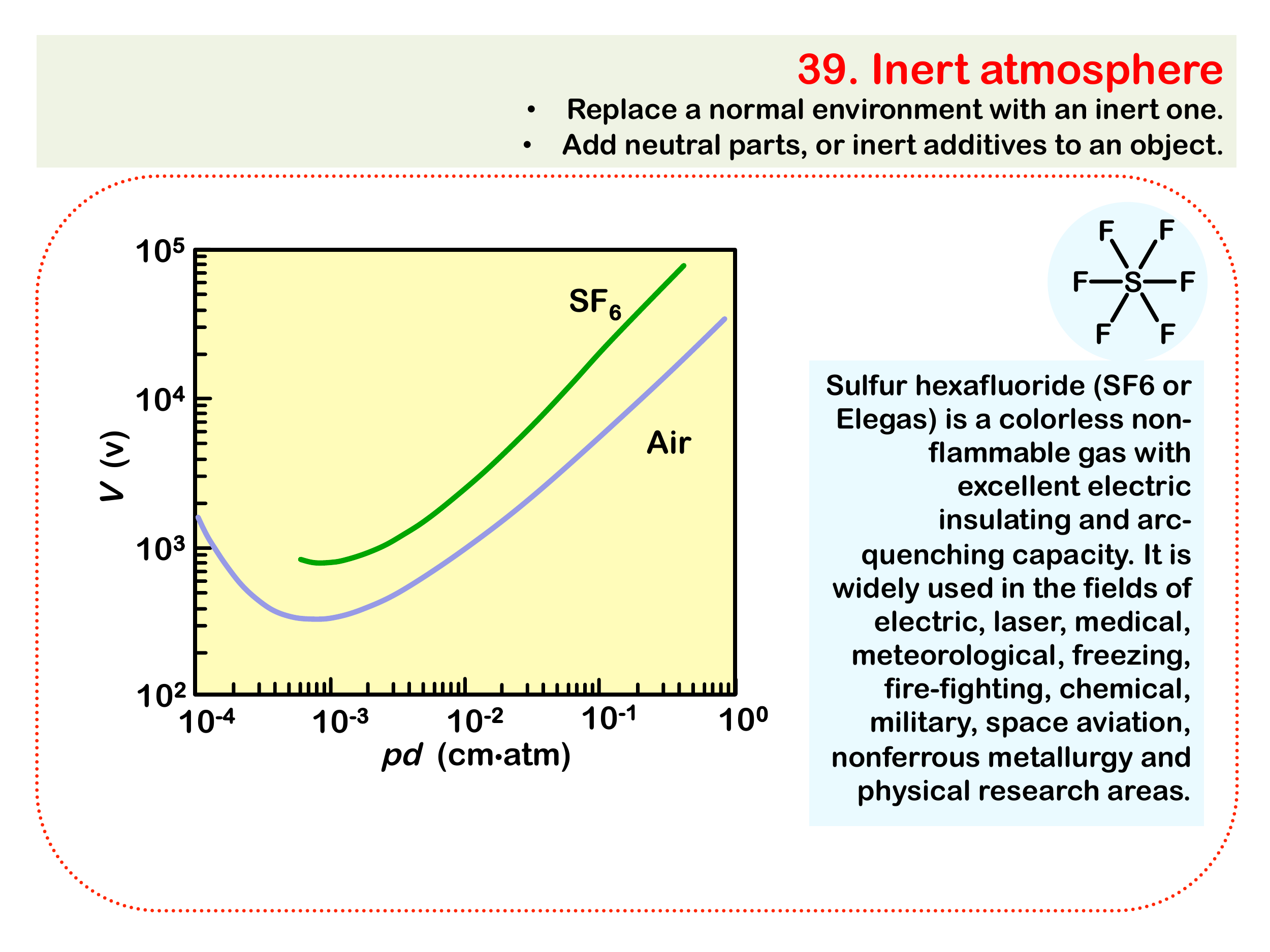}
\caption{Inventive principle ``Inert atmosphere''.}
\label{lab39}
\end{figure}

An example we selected to illustrate this principle is a method of using sulfur hexafluoride (SF6 or Elegas), which is a colorless non-flammable gas with excellent electric insulating and arc-quenching capacity. This gas can be used, in particular, to fill the interior of electrostatic generators (Van der Graaf or Cockcroft-Walton types) in order to reach higher voltage and therefore higher energy of accelerated particles. The curves on the left side of the picture show a comparison of electrical discharge voltage versus the value of pressure multiplied by the gap between electrodes for SF6 in comparison with an air.

\newpage
\section{Composite materials}

The inventive principle {\it composite materials} may involve changing from uniform to composite (multiple) materials. 

\begin{figure}[!h]
\hspace*{-1pc}
\includegraphics[width=1.05\textwidth]{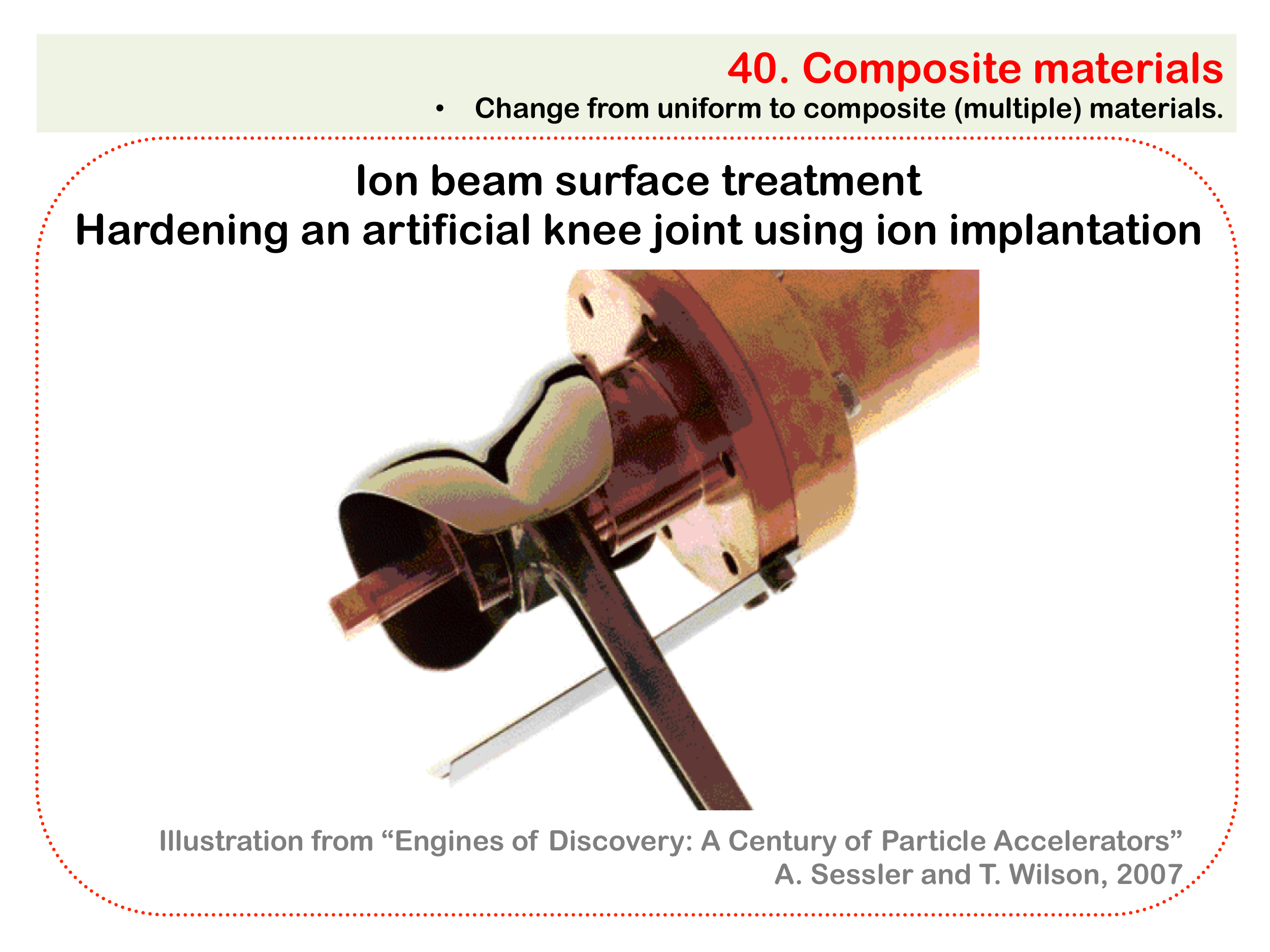}
\caption{Inventive principle ``Composite materials''.}
\label{lab40}
\end{figure}

An example we selected to illustrate this principle is a method of hardening of an artificial knee joint using ion implantation\footnote {A. Sessler and T. Wilson, {\it Engines of Discovery: A Century of Particle Accelerators}, 2007.}. In this case ion beam surface treatment results in a creation of a strong film on the surface of the artificial joint, equivalent to creating a composite material.

\newpage

\section*{Discussion and conclusion}

Suggested illustrations of inventive principles, we hope, will be useful to stimulate the readers to think about inventive ways to solve scientific problems they encounter in their research. The examples we suggested are often not ideal and could certainly be improved -- we welcome our readers to participate in creation of better examples. And most importantly, we welcome the readers to use the methodology of inventive problem solving in their research, and beyond.



\begin{thebibliography}{9}   

\bibitem{Altshuller-book}
Altshuller, G., ``Innovation Algorithm: TRIZ, systematic innovation and technical
creativity'', first ed. Worcester, MA: Technical Innovation Center, Inc. (1999).

\bibitem{Seryi-book}
Seryi, A., ``Unifying Physics of Accelerators, Lasers and Plasma'', CRC Press, Taylor \& Francis Group,  (2015).

\bibitem{Serye-book2}
Seryi, A.A., Seraia, E.I., ``Inventing Instruments of Future Science'', URSS Publishing company. In Russian. Published in June 2016, ISBN 978-5-9710-3185-7, (2016). 

\end{thebibliography}
\end{document}